\documentclass[11pt,a4paper]{article}

\usepackage[T1]{fontenc}
\usepackage[utf8]{inputenc}
\usepackage{lmodern}
\usepackage{microtype}
\usepackage{amsmath,amssymb,mathtools,bm}
\usepackage{booktabs}
\usepackage{array}
\usepackage{graphicx}
\usepackage{subcaption}
\usepackage{xcolor}
\usepackage{geometry}
\usepackage{hyperref}
\usepackage[nameinlink]{cleveref}
\usepackage[title,titletoc]{appendix}
\crefname{figure}{Fig.}{Figs.}
\Crefname{figure}{Figure}{Figures}
\crefname{section}{Sec.}{Secs.}
\Crefname{section}{Section}{Sections}
\crefname{subsection}{Sec.}{Secs.}
\Crefname{subsection}{Section}{Sections}
\crefname{subsubsection}{Sec.}{Secs.}
\Crefname{subsubsection}{Section}{Sections}
\crefname{equation}{Eq.}{Eqs.}
\Crefname{equation}{Equation}{Equations}
\crefname{appendix}{Appendix}{Appendices}
\Crefname{appendix}{Appendix}{Appendices}
\usepackage[numbers,sort&compress]{natbib}
\usepackage{enumitem}

\graphicspath{{figures/}}

\hypersetup{
  colorlinks=true,
  linkcolor=blue!45!black,
  citecolor=blue!45!black,
  urlcolor=blue!55!black
}

\newif\ifworknotes
\worknotesfalse

\newcommand{\E}{\mathbb E}
\newcommand{\Pp}{\mathbb P}
\newcommand{\Var}{\operatorname{Var}}

\newcommand{\diag}{\operatorname{diag}}
\newcommand{\Dord}{D_{\rm ord}}
\newcommand{\Dfix}{D_{\rm fix}}
\newcommand{\dcond}{\delta_{\rm cond}}
\newcommand{\muhat}{\widehat\mu}
\newcommand{\phione}{\phi(1)}
\newcommand{\Vstat}{V_{\rm stat}}
\newcommand{\trans}{\mathsf T}

\title{Coverage Is Not Ordering:\\
Ancillary Leakage and Representation Dependence in Covariance-Based Inference}

\author{
Tommaso Dorigo\\[0.5em]
\small INFN, Sezione di Padova, Padova, Italy;\\
\small Department of Computer Science, Electrical and Space Engineering,\\
\small Luleå University of Technology, Luleå, Sweden
}
\date{\today}

\begin{document}
\maketitle

\begin{abstract}
A covariance matrix is often used as a compact surrogate for the statistical model of a measurement. We show that this replacement can change not only a fitted value or its uncertainty, but also the likelihood-ratio ordering of the experiment itself. In a tractable correlated-measurement model, the data separate into an informative component and an ancillary residual disagreement. Exact likelihood inference therefore does not use that disagreement, whereas data-dependent covariance matrices and Gaussian reconstructions after nonlinear transformations can reintroduce it into parameter inference.

We quantify the resulting change by the probability mass of the symmetric difference of equally calibrated acceptance regions, \(D_{\rm ord}\). Operationally, \(D_{\rm ord}\) is the fraction of repeated experiments for which the confidence decision about a tested parameter value changes. In the small-uncertainty regime the ordering discrepancy is generically first order in the total relative uncertainty, whereas conventional and ancillary-conditioned coverage defects begin at second order. At “coverage-blind” points the leading coverage difference vanishes while the ordering discrepancy remains nonzero. In a representative blind case with 10\% total relative uncertainty, about 7.6\% of experiments change their confidence decision despite exact calibration of both procedures.

The same mechanism connects Peelle’s Pertinent Puzzle to the classical literature on ancillarity and relevant subsets: a covariance approximation can manufacture inferential relevance for a goodness-of-fit statistic that is ancillary in the true model. We show directly how this changes the confidence interval reported for the same informative content, and extend the analysis beyond equal statistical uncertainties and beyond the two-measurement case.

\end{abstract}

\section{Introduction}
\label{sec:intro}



Experimental measurements in fundamental science are often reported by quoting central values and uncertainties around them; when several quantities are measured jointly, a covariance matrix is usually provided as well.  These compact summaries may become essential inputs to subsequent combinations and reinterpretations, even though they contain only part of the information in the original measurement results. This problem is well-known and widely discussed, and specific prescriptions have been proposed to avert it.  The question we wish to address in this work is which inferential properties are robust to the replacement of that full model by a covariance-based Gaussian reconstruction.

Two familiar situations make the issue worth the attention of experimentalists. One is Peelle’s Pertinent Puzzle (PPP), a name that has been applied in the literature to a family of closely related anomalies in strongly correlated combinations.  Here we focus on the particularly important case
in which a shared multiplicative uncertainty is converted into an absolute
covariance using the observed central values themselves.
A downward fluctuation then also reduces the covariance assigned to that observation, so that the same fluctuation can acquire excessive influence in the fit. The resulting combination is systematically pulled downward and can, in
sufficiently discrepant cases, even lie below the individual measurements\footnote{It is important to clarify in passing that the above mechanism should be distinguished from the familiar possibility of negative
weights in a BLUE combination with a strongly correlated but fixed covariance
matrix \cite{LyonsGibautClifford1988,Nisius2014,ValassiChierici2014,BurrEtAl2011};
such weights can place the combined estimate outside the range of the input
measurements without, by themselves, implying the data-dependent bias
mechanism studied here.
}
\cite{LyonsGibautClifford1988,Dagostini1994,ChibaSmith1994,
HansonKawanoTalou2005,NeudeckerFruehwirthLeeb2012,BallEtAl2010,
WaltonEtAl2025}.
Neudecker, Fr\"uhwirth and Leeb made this point particularly explicit.  In a
Bayesian treatment of measurements sharing a common stochastic normalization,
they showed that the anomalous PPP mean results from using the individually
fluctuating observations to construct the normalization covariance, rather
than from nonlinearity alone.  Replacing those individual factors by a common
estimate of the underlying quantity restores the correct posterior mean and,
with the corresponding variance prescription, the correct first two posterior
moments \cite{NeudeckerFruehwirthLeeb2012}. The same covariance diagnosis was subsequently tested beyond a single
physical quantity and applied to a nuclear-data evaluation
\cite{NeudeckerEtAl2014}. Our starting point is consistent
with that diagnosis, but the question pursued here goes beyond the fitted
moments: once such a data-dependent covariance is regarded as an approximate
likelihood, what does it do to likelihood-ratio ordering, ancillary dependence,
and confidence decisions even after coverage has been separately calibrated? The issue remains current in high-energy physics (HEP): a recent neutrino-interaction study examined PPP using MicroBooNE and T2K measurements and advocated model-fitting exercises as a routine diagnostic
in cross-section publications \cite{AbeEtAl2026}.

A second, related issue arises when the same measurement is
expressed in a nonlinear variable. In HEP practice there are many situations when this occurs: for example, a cross section may be
converted into a coupling strength, and a measured lifetime is often the basis for reporting a particle decay width.  When the full likelihood is available, such one-to-one changes of coordinates cannot alter likelihood-ratio inference; instead, a covariance propagated to the new variable and
used to reconstruct a Gaussian likelihood need not share that invariance
\cite{ChibaSmith1994,BoxCox1964}.  

We show that the two mentioned phenomena are manifestations of a common mechanism: covariance reconstruction can change which features of the observed data influence the inference.
In order to understand the problem and unearth its generating mechanism, we study in this work the smallest correlated-measurement model in which the mentioned effects can be separated analytically.  The model has a common mean, ordinary Gaussian measurement errors, and one shared multiplicative random normalization.  Its simplicity allows us to identify exactly which part of the data contains information about the parameter, and therefore to see when an approximate covariance treatment introduces an inferential dependence that is
absent from the original statistical model.

We compare the exact likelihood with two covariance-level reconstructions
that commonly arise when experimental results are reused.  In the first, a
multiplicative uncertainty is encoded through a covariance evaluated from the
observed data.  In the second, the measurement is expressed in a nonlinear
variable and a Gaussian likelihood is rebuilt from the transformed covariance.
These examples will allow us to check whether a compact covariance description
preserves not only fitted values and uncertainties, but also the way the
original measurement ranks possible experimental outcomes.

\paragraph{Why ordering matters.}
The Neyman construction, which sits at the foundations of frequentist-based inference and is commonly used in HEP measurements, fixes the probability content of an acceptance region, but by itself does not prescribe which particular outcomes should constitute that region. An additional prescription is therefore needed to select which outcomes enter the acceptance region; this is commonly expressed through an ordering rule, and infinitely many such choices are possible.
Of special relevance to HEP practice is likelihood-ratio ordering, which became especially familiar through the unified construction of Feldman and Cousins \cite{FeldmanCousins1998}. We take that ordering as our model-based reference and ask whether a covariance representation intended to approximate the same measurement preserves it. 
In doing that we do not claim that any change of ordering following the representation change is intrinsically pathological; yet the mechanism must be understood, because the resulting confidence decisions can change in ways that are invisible to coverage alone. Once likelihood-ratio ordering has been adopted because it ranks outcomes according to their relative compatibility with the tested model, a representation-dependent change of that ordering means that the approximation has altered the very criterion by which experimental outcomes are judged.

This distinction is important.  Suppose that two confidence constructions
are each calibrated to accept $68.27\%$ of repeated experiments.  Equal
coverage tells us that they accept the same amount of probability; yet it
does not tell us that they accept the same experiments --and in fact they need not do so.  A data set may lie inside one acceptance region and outside the other even when both procedures have exactly the advertised coverage.  After inversion, this is not merely a geometrical difference between two acceptance regions: the procedures make different statements about whether the tested parameter value belongs to the reported confidence interval.

The likelihood-ratio ordering prescription employed here gives us a clean way to expose such a difference, provided it is understood that it is not the source of the effect we focus on: applied to the exact likelihood it inherits the information structure of the exact model.  Applied to an approximate likelihood, instead, it reveals whether the approximation has changed the ranking of possible experimental outcomes.

\paragraph {Main findings and consequences.}
The individual ingredients of our analysis have substantial precedents in
the literature: the pathologies of data-dependent covariance combinations,
the limitations of covariance propagation under nonlinear transformations,
likelihood-ratio confidence constructions, and the role of ancillary and
recognizable subsets are all well established.  To our knowledge, however,
their connection through ancillary leakage, and the resulting separation
between likelihood-ratio ordering and coverage developed below, has not been
established in this form. Our analysis leads to four main findings.

\begin{enumerate}[leftmargin=2.2em,label=(\roman*)]

\item
In the exact measurement model, the observed data decompose into a weighted mean \(m\), which carries the information about the common parameter \(\mu\), and a residual disagreement \(H\), which is ancillary for \(\mu\). The latter describes how strongly the individual measurements disagree, but its distribution contains no information about \(\mu\). A
data-dependent covariance can nevertheless make inference on the mean depend
on $H$; we call this \emph{ancillary leakage}.

\item
The covariance approximations studied here can change the likelihood-ratio ordering of the sample space.
To make this statement concrete, fix a value of the parameter $\mu$ and
consider all possible data sets that could be obtained in repeated
experiments.  These outcomes form the sample space of possible measurements; for the two-measurement example below, it is simply the $(x_1,x_2)$ plane.
Let $A_E(\mu)$ denote the subset of this space accepted by the
likelihood-ratio construction based on the exact likelihood, and
$A_A(\mu)$ the corresponding subset obtained from the approximate
likelihood.  
We calibrate the two regions separately so that each has the same probability content \(\gamma\) under the true model at \(\mu\).
By ``equally calibrated'' we mean that
\[
\Pp_\mu[X\in A_E(\mu)]
=
\Pp_\mu[X\in A_A(\mu)]
=
\gamma ,
\]
with $\gamma=0.6827$ in the examples below.

We quantify their disagreement by
\begin{equation}
\Dord(\mu)=
\Pp_\mu\!\left[
A_A(\mu)\triangle A_E(\mu)
\right],
\label{eq:intro_dord}
\end{equation}
where $\triangle$ denotes the symmetric difference: the set of experimental
outcomes accepted by one construction and rejected by the other.
Thus $\Dord$ has a direct operational meaning.  It is the fraction of
repeated experiments for which the exact and approximate confidence
constructions make different include/exclude decisions about $\mu$.

The size of the effect quantified by $\Dord$ is especially revealing in the small-uncertainty
regime.  Writing $\lambda$ for the total relative uncertainty scale, we find
generically
\begin{equation}
\boxed{
\Dord=O(\lambda),
\qquad
\text{coverage diagnostics considered here}=O(\lambda^2).
}
\label{eq:intro_hierarchy}
\end{equation}
The two coverage diagnostics studied below are the fixed-threshold coverage difference
$$ \Dfix \equiv P_\mu(q_A\le 1)-P_\mu(q_E\le 1), $$
which compares the exact and approximate likelihood-ratio statistics at the same conventional threshold, and the ancillary-conditioned defect
$$ \dcond(H)\equiv P_\mu(A_A\mid H)-\gamma, $$
which measures the departure from nominal coverage after conditioning on the observed disagreement \(H\).
Neither quantity is the unconditional coverage difference of the separately calibrated belts, which is zero by construction.

The point of the hierarchy is that experiments
can begin to move between the accepted and rejected regions at first order in
the uncertainty, whereas the net probability imbalance seen by these coverage
checks appears only at second order.  Ordering is therefore a parametrically
more sensitive diagnostic of the approximation.

\item
There are \emph{coverage-blind} loci where the leading fixed-threshold
coverage difference vanishes while the first-order ordering discrepancy
remains nonzero.  This occurs not only for nonlinear covariance
reconstructions in the two-measurement problem, but already for the
data-centered PPP construction, in which the multiplicative covariance is
evaluated from the observed central values, once \(N\ge3\) measurements are
combined.  At such points a conventional coverage comparison can look
essentially perfect while a non-negligible fraction of experiments lead to
different confidence decisions.

\item
For the data-centered PPP prescription the ordering problem can be solved
analytically for arbitrary finite \(N\) and arbitrary statistical variances. The entire ancillary dependence continues to enter through the single goodness-of-fit statistic \(H\). At leading order, the ordering discrepancy grows linearly with the overall uncertainty scale, proportionally to the normalization contribution and to the \(N-1\) residual degrees of freedom. For every \(N\geq 3\), there is moreover a physically allowed parameter choice at which the leading coverage difference vanishes while the ordering discrepancy remains nonzero. Thus neither the ancillary leakage nor the ordering–coverage separation is an artifact of the equal-error, two-measurement geometry.

\end{enumerate}

\paragraph{Consequences for experimental practice.}
The central distinction our work revolves around is that 
coverage controls how much probability is accepted, while ordering specifies
which experiments make up that probability.  In the model studied below we
will exhibit data sets with the same informative weighted mean but different
ancillary disagreement.  The exact likelihood assigns them the same
likelihood-ratio ordering, and after inversion the same confidence interval; yet
the covariance approximation does not. This is particularly striking in a coverage-blind example we provide in \cref{sec:blind}, where the
exact and approximate procedures are both calibrated to $68.27\%$, yet
approximately $7.65\%$ of repeated experiments change their confidence
decision about the true parameter value.  The discrepancy therefore survives
even after coverage has been made exactly equal: what changes is which
experiments contribute to that coverage.

The ancillary interpretation also connects the problem to the classical
literature on ancillarity, recognizable subsets, and relevant subsets
\cite{Fisher1956,Cox1958,Buehler1959,Wallace1959,Kiefer1977,
Robinson1979,GoutisCasella1995,Cousins2011,GhoshReidFraser2010,
MullerNorets2016,PawitanLeeLee2023}.  Cousins emphasized, for a different
high-energy-physics interval problem, that exact unconditional coverage can
conceal nontrivial behavior in recognizable subsets \cite{Cousins2011}. In our model the recognizable quantity is especially concrete: it is simply the internal
disagreement of two measurements.  In the exact model that disagreement is
ancillary for the parameter of interest; the approximation is what gives it
inferential relevance.

\paragraph{Plan of the paper.}
We have organized the paper around the sequence of questions an experimental
physicist might naturally ask.  \Cref{sec:model} first asks
which part of the observed data actually contains information about the common
mean, and answers this by separating the informative weighted mean from the
ancillary disagreement.
In \Cref{sec:covariance} we then ask how a covariance approximation can make
that disagreement matter, and why the answer can depend on the variable in
which the result is expressed.  \Cref{sec:ordering} addresses the next
question: whether two procedures calibrated to the same coverage can
nevertheless rank experiments differently.  We quantify the resulting
ordering mismatch, derive its leading behavior, and show why conventional
coverage diagnostics can respond only at higher order or even become
locally blind to it.

\Cref{sec:relevant} examines what unconditional coverage can hide when one
conditions on the ancillary disagreement, while \cref{sec:intervals} shows
how the effect propagates to the confidence interval actually reported.
Finally, \cref{sec:robustness,sec:practice} test the robustness of the
mechanism and draw its implications for experimental practice.

\section{The measurement problem and its hidden ancillary}
\label{sec:model}

We begin with the simplest measurement model that contains the ingredients
needed for the problem studied in this paper.  Two experimental determinations
measure the same positive quantity $\mu$, each with its own statistical
uncertainty, while sharing a common relative normalization uncertainty:
\begin{equation}
X_i=\mu(1+s\xi)+\epsilon_i,
\qquad i=1,2.
\label{eq:model}
\end{equation}
Here
\begin{equation}
\xi\sim\mathcal N(0,1),
\qquad
\epsilon_i\sim\mathcal N(0,\sigma_i^2),
\end{equation}
and all random variables are independent.  The number $s$ is the fractional
size of the shared normalization uncertainty, while $\xi$ is the latent
standard-normal variable used to represent it.  Thus a given repetition of
the experiment has a common relative offset $s\xi$, in addition to the
independent statistical errors $\epsilon_i$.

Before constructing any likelihood approximation, it is useful to ask a
basic question: which combinations of the two observed numbers actually
contain information about $\mu$?  There are two natural pieces of the data.
One tells us where the pair lies along the common-mean direction; the other
tells us how strongly the two measurements disagree with one another.

Define the inverse-variance weights
\begin{equation}
w_i=\sigma_i^{-2},
\qquad
\omega=w_1+w_2,
\qquad
\tau^2=\omega^{-1},
\end{equation}
and the usual weighted mean
\begin{equation}
m=\frac{w_1x_1+w_2x_2}{\omega}.
\label{eq:m}
\end{equation}
This is the coordinate of the observed pair along the direction corresponding
to a common value of the two measurements.

The remaining information is the disagreement between the measurements.
It is useful first to write it as the signed standardized contrast
\begin{equation}
U=
\frac{x_1-x_2}
{\sqrt{\sigma_1^2+\sigma_2^2}}.
\label{eq:U}
\end{equation}
The common normalization uncertainty cancels from $x_1-x_2$, so $U$ measures
only the internal statistical disagreement of the two measurements.  Its
square is
\begin{equation}
H\equiv U^2
=
\sum_i w_i(x_i-m)^2
=
\frac{(x_1-x_2)^2}{\sigma_1^2+\sigma_2^2}.
\label{eq:H}
\end{equation}
Thus $H$ is simply the usual one-degree-of-freedom chi-square measuring the
compatibility of two determinations of a common quantity.

The terminology ``residual'' and ``orthogonal'' can also be given a direct
geometrical meaning.  Writing
\begin{equation}
x=m\mathbf1+r,
\qquad
\mathbf1=(1,1)^\trans,
\end{equation}
the residual vector $r$ is what remains after subtracting the common weighted
mean.  In the inverse-variance metric
$W=\diag(w_1,w_2)$ it obeys
\begin{equation}
\mathbf1^\trans Wr=0.
\end{equation}
The mean and residual directions are therefore orthogonal in precisely the
metric used to combine the two measurements.

The Gaussian model makes the statistical separation equally simple.
Because the common normalization term cancels from the difference,
\begin{equation}
U\sim\mathcal N(0,1),
\end{equation}
and hence
\begin{equation}
H=U^2\sim\chi^2_1.
\end{equation}
Moreover, the weighted statistical error entering $m$ is uncorrelated with
the contrast $U$.  Since the variables are jointly Gaussian, this implies
statistical independence.  Consequently,
\begin{equation}
m\mid\mu\sim
\mathcal N\!\left(\mu,\tau^2+s^2\mu^2\right),
\qquad
H\sim\chi^2_1,
\qquad
m\ \hbox{is independent of}\ H.
\label{eq:factorization}
\end{equation}
Equivalently,
\begin{equation}
\boxed{
f(m,H\mid\mu)=f(m\mid\mu)\,f(H).
}
\label{eq:factorization_pdf}
\end{equation}

This factorization is the reference point for everything that follows.
All dependence of the exact sampling distribution on the parameter $\mu$ is
contained in $m$.  The statistic $H$ tells us whether the two measurements
agree unusually well or unusually poorly, but its distribution is completely
independent of $\mu$.  In this sense $H$ is ancillary for the parameter of
interest.

This observation gives us a particularly clean diagnostic.  If an
approximation to the same measurement later makes the inference on $\mu$
depend on $H$, that dependence cannot have come from information about $\mu$
present in the original model: it has been introduced by the approximation.
This is the mechanism we will refer to as \emph{ancillary leakage}.

\begin{figure}[t]
\centering
\includegraphics[width=0.7\linewidth]{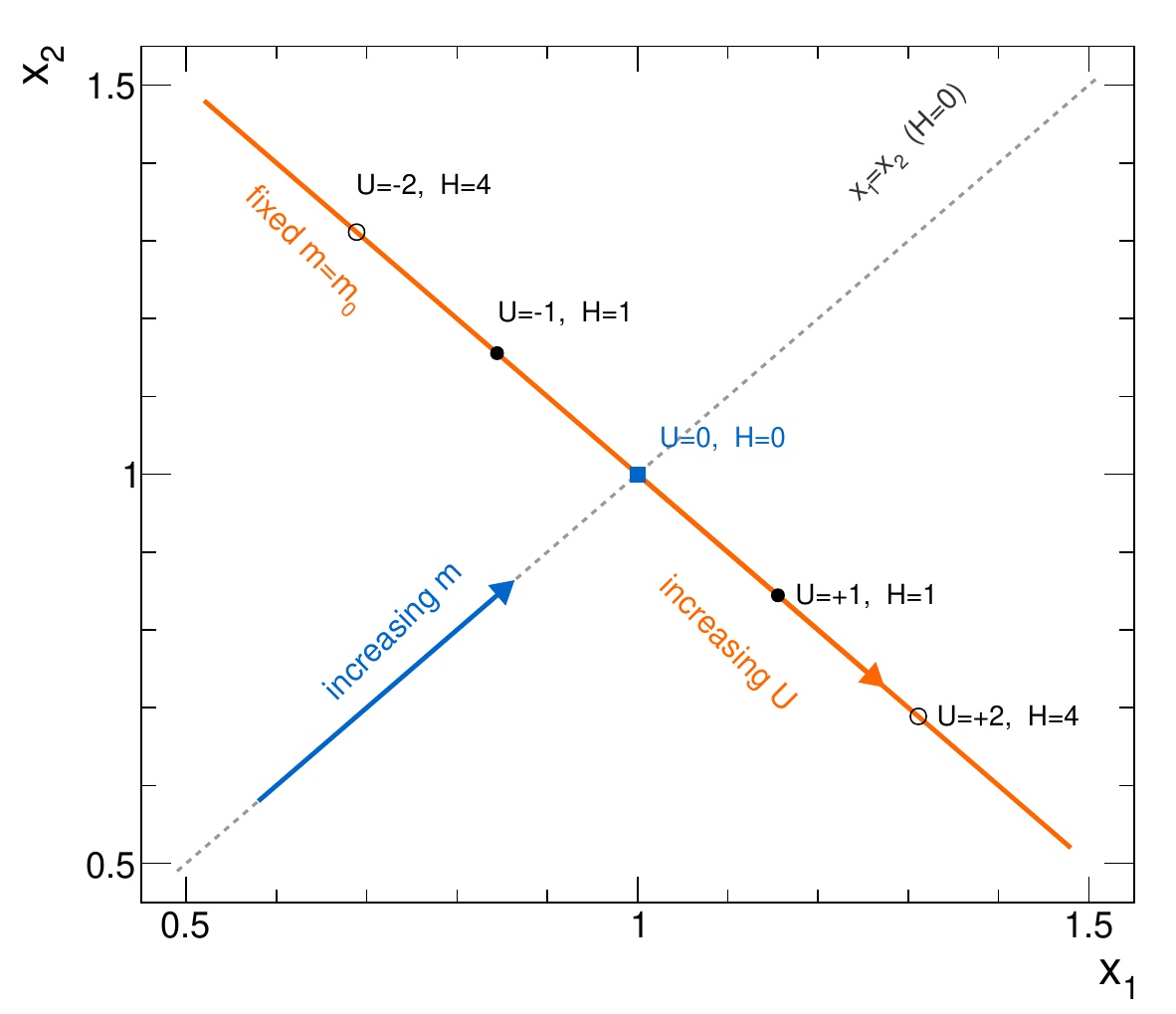}
\caption{Informative and ancillary directions in the two-measurement
experiment, where measurements $x_1$ and $x_2$ have identical uncertainties $\sigma_1=\sigma_2$. Moving along a line of fixed $m$ changes the internal
disagreement $H$ without changing the exact likelihood for $\mu$. See the text for detail.}
\label{fig:mHgeometry}
\end{figure}

\subsection{Three statistical descriptions of the same uncertainty \label{s:3descr}}

The phrase ``a common normalization uncertainty of size $s$'' specifies a
physical uncertainty model, but does not by itself determine how that model
will be represented in an inference procedure.  Before studying ordering, it
is therefore useful to distinguish three statistical objects that recur in
practice.

If the Gaussian normalization variable $\xi$ is treated as the random effect
appearing in \cref{eq:model} and integrated out, the exact marginal
likelihood is
\begin{equation}
-2\log L_{\rm marg}(\mu)
=
\frac{\omega(m-\mu)^2}{1+\omega s^2\mu^2}
+\log(1+\omega s^2\mu^2)+\mathrm{const}.
\label{eq:marginal}
\end{equation}
As anticipated by the factorization above, the marginal likelihood depends on $m$ but not on $H$.

A closely related construction treats the common normalization as a nuisance
parameter constrained by its Gaussian uncertainty and profiles over it.  The
resulting quadratic function is
\begin{equation}
\chi^2_{\rm prof}(\mu)
=
H+\frac{\omega(m-\mu)^2}{1+\omega s^2\mu^2}.
\label{eq:profile}
\end{equation}
Here $H$ appears explicitly, but only as an additive term independent of
$\mu$: it therefore cancels from likelihood ratios.  Both constructions
preserve the basic separation found above: the internal disagreement does not
affect the inference on the common mean.

In what follows we use the marginal likelihood as our exact reference;
the profiled form is included here to emphasize that the ancillary
contribution \(H\) cancels from likelihood ratios in either treatment.

The situation changes when the multiplicative uncertainty is instead encoded
through a covariance evaluated from the observed central values---a
prescription we will refer to as \emph{data-centered}:
\begin{equation}
\widehat V_x=\Vstat+s^2xx^\trans,
\qquad
\Vstat=\diag(\sigma_1^2,\sigma_2^2),
\label{eq:plugin_cov}
\end{equation}
where $x=(x_1,x_2)^\trans$.  Now the covariance used to judge the data is
itself built from the observed values.  The associated quadratic form has its
minimum at
\begin{equation}
\boxed{
\muhat_{{\rm dat},1}=\frac{m}{1+s^2H}.
}
\label{eq:plugin_est}
\end{equation}
This is the same PPP displacement exhibited, in equivalent notation, by
Neudecker, Fr\"uhwirth and Leeb \cite{NeudeckerFruehwirthLeeb2012}.  The
form above makes explicit that the displacement is controlled by the ancillary
goodness-of-fit statistic $H$.
The contrast with the exact model is immediate: an ancillary goodness-of-fit
statistic now changes the estimate of the parameter.  The larger the internal
disagreement $H$, the farther the estimate is pulled below the weighted mean
$m$.

We can examine in detail the situation by considering a simple numerical example:
\begin{equation}
x_1=10,\quad x_2=11,\quad
\sigma_1=\sigma_2=0.5,\quad s=0.2,
\end{equation}
where one has $m=10.5$, $H=2$, and
\begin{equation}
\muhat_{{\rm dat},1}=9.7222.
\end{equation}
The exact marginal and profiled constructions have no analogous
$H$-dependent displacement.  The point is not that an estimate outside the
range of its inputs is intrinsically wrong---fixed-covariance BLUE
combinations can legitimately have negative weights
\cite{Nisius2014,ValassiChierici2014,BurrEtAl2011}---but that in the present construction
the metric used for inference is itself determined by the fluctuating data.

We have therefore identified the first effect of the covariance
approximation: it can make a quantity that is ancillary in the exact model
influence the fitted common mean.  This is already enough to expose the PPP
mechanism at the point-estimate level.  The next question is whether the
problem is tied only to this particular representation of the measured
quantity.  In \cref{sec:covariance} we ask what happens when the same
measurement is expressed in a nonlinear variable and a new Gaussian
likelihood is reconstructed from the transformed covariance.

\section{From covariance approximation to representation dependence}
\label{sec:covariance}

The PPP estimator already demonstrates ancillary leakage at the point-estimate
level.  A second question appears when published measurements are reused in a
different representation.
A physicist may, for example, convert a rate to a coupling, with known
proportionality factor $K$,
\begin{equation}
\sigma_{\rm xs}=K g^2,
\qquad
g=\sqrt{\sigma_{\rm xs}/K},
\end{equation}
or a lifetime $\tau_{\rm life}$ to a width,
$\Gamma=\hbar/\tau_{\rm life}$.  With the full likelihood these
are harmless one-to-one changes of coordinates.  If $y=h(x)$,
\begin{equation}
f_Y(y\mid\mu)=
f_X(h^{-1}(y)\mid\mu)
\left|\det\frac{\partial h^{-1}}{\partial y}\right|,
\end{equation}
and the data-only Jacobian cancels from likelihood ratios.

A propagated covariance,
\begin{equation}
V_y\simeq J V_xJ^\trans,
\label{eq:delta_method}
\end{equation}
is instead only a local approximation to the transformed statistical model.
At this point one must specify where the Jacobian used for the covariance
propagation is evaluated.  We distinguish two prescriptions.  In a
\emph{model-centered} reconstruction, the transformation is linearized at
the value predicted by the candidate model, $x_i=\mu$, so that the relevant
Jacobian is $J_\mu$.  In a \emph{data-centered} reconstruction, it is
linearized at the observed central values $x_i$, giving a Jacobian $J_x$.
The transformation $h$ is the same in the two cases; what differs is the
point about which its local linear approximation is made.
This distinction matters because an exact change of variable and a Gaussian
reconstruction treat the Jacobian in different ways.  In the exact
transformed likelihood, the Jacobian of the transformation is evaluated at
the observed data.  Once the data have been observed, this factor is
independent of the candidate value of $\mu$, and therefore cancels from a
likelihood ratio.  In a model-centered covariance reconstruction, by contrast, the uncertainties are propagated using the derivative of the transformation evaluated at the candidate model prediction, \(h'(\mu)\). For a nonlinear transformation this derivative depends on \(\mu\), so the reconstructed covariance changes as the candidate parameter value is varied.

Consider first the model-centered construction.  Two effects then appear.
First, the transformed residual
$h(x)-h(\mu)$ is not exactly equal to the linearized residual
$h'(\mu)(x-\mu)$; this produces a curvature correction to the quadratic
part of the likelihood.  Second, the propagated covariance changes its
volume as $h'(\mu)$ changes.  Since a normalized Gaussian contains the
factor $1/\sqrt{\det V}$ as well as the usual quadratic exponent, this
produces an additional parameter-dependent contribution.
The difficulty of transporting covariance information through nonlinear derived quantities had already been emphasized by Zhao and Perey, who noted
that such covariance matrices are approximate because unknown true
quantities are replaced by their measured estimates, and that different evaluation strategies can consequently become inconsistent \cite{ZhaoPerey1992}.

Chiba and Smith showed that the quadratic least-squares term can remain invariant when residuals, covariance, and sensitivities are transformed
consistently at the same linear order \cite{ChibaSmith1994}.  That result
does not, however, guarantee invariance of the full normalized Gaussian
likelihood.  Up to terms that depend only on the observed data, the
difference from the exact transformed likelihood can be written
schematically as
\begin{equation}
\boxed{
\Delta_h(\mu;x)
=
\underbrace{\Delta Q_{\rm curvature}}_{\text{nonlinear residual}}
+
\underbrace{2\log|\det J_\mu|}_{\text{Gaussian volume}}.
}
\label{eq:Deltah}
\end{equation}
The second term is independent of $\mu$ only when the transformation has a
constant slope, $h(x)=ax+b$.  For a genuinely nonlinear transformation the
local stretching of the coordinate system changes with $\mu$, and the
normalization of the reconstructed Gaussian therefore changes the likelihood
ratio itself.  For example, $h'(\mu)\propto\mu^{-1/2}$ for a square-root
transformation and $h'(\mu)=1/\mu$ for a logarithm, so in both cases the
Gaussian-volume term varies explicitly across the likelihood scan.

The data-centered prescription behaves differently.  There the Jacobian is
evaluated at the observed values, so once a particular data set has been
observed it does not generate the same $\mu$-dependent Gaussian-volume term.
The price is instead that the metric used for inference now depends on the
realized data themselves.  Different statistical fluctuations therefore
produce different propagated covariances.  This is the same basic mechanism
already encountered in the PPP construction, now combined
with the nonlinear dependence of the transformed residuals.  Thus
model-centering and data-centering represent two distinct ways in which a
Gaussian reconstruction can depart from the invariance of the exact
likelihood.

To probe these effects continuously we use the Box--Cox family
\cite{BoxCox1964},
\begin{equation}
h_p(x)=
\begin{cases}
(x^p-1)/p,&p\neq0,\\
\log x,&p=0,
\end{cases}
\label{eq:boxcox}
\end{equation}
with three representative cases:
\begin{equation}
p=1,\qquad p=\tfrac12,\qquad p=0.
\end{equation}
The case \(p=1\) gives \(h_1(x)=x-1\), which differs from the original
variable only by an irrelevant constant shift; we therefore refer to it as
the untransformed case.  The choices \(p=\tfrac12\) and \(p=0\) correspond
respectively to a square-root-like coupling transformation and the logarithm.

The explicit transformed likelihoods for the model-centered and
data-centered prescriptions are collected in \cref{app:transform}.  The
important point for the main argument is simple: replacing the exact
transformed likelihood by a covariance-based Gaussian reconstruction changes
the statistical model.  It can therefore change the ordering of possible
data sets even when the transformation itself is one-to-one and the original
and transformed central values look innocuous.

\section{Ordering mismatch despite exact coverage}
\label{sec:ordering}

For a tested value $\mu$, a Neyman construction selects an acceptance region
$A_\mu$ with the desired probability.  We use subscripts $E$ and $A$ for the
exact and approximate constructions, respectively.
For either construction we define the likelihood-ratio statistic
\[
q(\mu;x)\equiv -2\log\frac{L(\mu;x)}{L(\hat\mu;x)},
\]
and use \(q_E\) and \(q_A\) for the exact and approximate
likelihoods. Feldman--Cousins
likelihood-ratio ordering \cite{FeldmanCousins1998} supplies a principled ranking of the sample
space when the likelihood is the true model.

The exact marginal likelihood ratio depends only on $m$.  For the local
ordering calculations we now specialize to equal statistical errors and scale
all measurements by the true value $\mu_0$, so that $\mu_0=1$.  We keep the
same symbols $x_i$ and $m$ for the resulting dimensionless quantities and
write
\begin{equation}
m=1+\lambda Z,
\qquad
x_{1,2}=1+\lambda Z\pm\rho U,
\qquad
Z,U\stackrel{\rm iid}{\sim}\mathcal N(0,1).
\end{equation}
Here
\begin{equation}
\rho=\frac{\tau}{\mu_0},
\qquad
\lambda^2=s^2+\rho^2,
\qquad
\alpha=\frac{\rho^2}{\lambda^2},
\qquad
\beta=\frac{s^2}{\lambda^2}=1-\alpha,
\qquad
H=U^2.
\label{eq:lambdalpha}
\end{equation}
Thus $\lambda$ is the total relative uncertainty, while $\alpha$ and $\beta$
are respectively its statistical and normalization fractions.  
Throughout this section, \(\phi\) and \(\Phi\) denote the standard
normal density and cumulative distribution function, and
\begin{equation}
\gamma\equiv 2\Phi(1)-1=0.682689\ldots
\label{eq:gamma}
\end{equation}
is the nominal coverage used in the numerical examples.

The exact FC belt is independent of $H$. Conversely, a covariance-based approximate likelihood will generally produce boundaries that bend with $H$.  This can be seen most easily before doing any expansion.

\begin{figure}[t]
\centering
\includegraphics[page=2,width=0.98\linewidth]{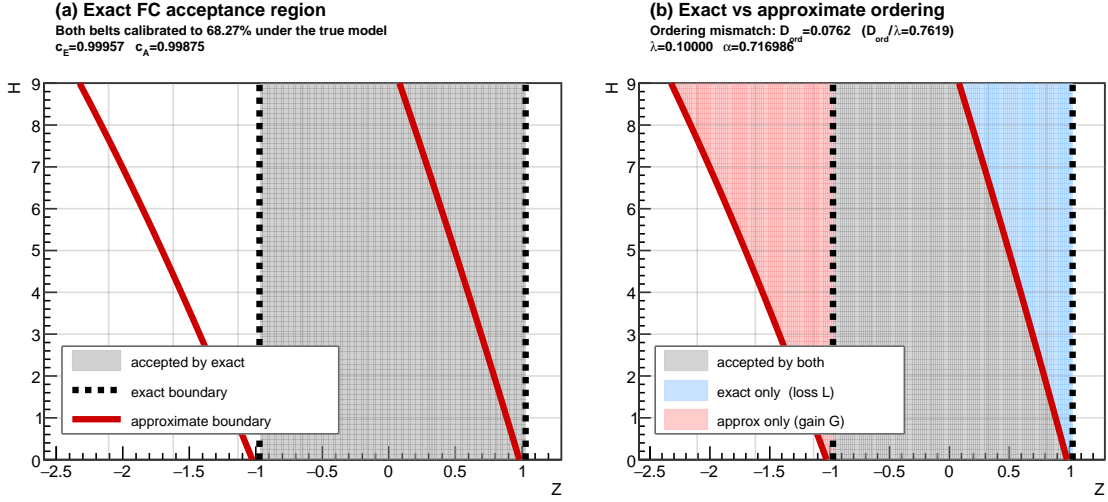}
\caption{Exact and approximate likelihood-ratio ordering for the data-centered logarithmic reconstruction at a representative parameter point. Both constructions are calibrated to the same \(68.27\%\) probability under the true model. The exact acceptance region is independent of the ancillary \(H\); the approximate one is not. Blue and red regions exchange confidence decisions while preserving the total accepted probability. The parameter point shown here is the coverage-blind case analyzed later in \cref{sec:blind}.}
\label{fig:belts}
\end{figure}

To quantify the discrepancy between the two acceptance regions, define
\begin{equation}
\boxed{
\Dord(\mu)
=
\Pp_\mu\!\left[
A_A(\mu)\triangle A_E(\mu)
\right].
}
\label{eq:dord}
\end{equation}
Here $\triangle$ denotes the symmetric difference of the two
acceptance regions.  Explicitly,
\begin{equation}
A_A\triangle A_E
=
(A_A\setminus A_E)\cup(A_E\setminus A_A).
\label{eq:symmetric_difference}
\end{equation}
The symbol $\setminus$ denotes set subtraction: $A_A\setminus A_E$ is the
set of experimental outcomes accepted by the approximate construction but
not by the exact one, whereas $A_E\setminus A_A$ contains those accepted by
the exact construction but not by the approximate one.  Thus the symmetric
difference is simply the union of the two exchanged regions shown in
\cref{fig:belts}.

The probability mass of a symmetric difference has been used in the literature as a natural measure of disagreement between acceptance (or, equivalently, rejection)
regions \cite{BohringerLohmann2022}.  Here its finite-sample interpretation
is especially direct.  If both belts have the same coverage, the probability
lost from the exact acceptance region must equal the probability gained by
the approximate one:
\begin{equation}
\Pp_\mu\!\left(
\underbrace{A_E\setminus A_A}_{\text{accepted only by exact}}
\right)
=
\Pp_\mu\!\left(
\underbrace{A_A\setminus A_E}_{\text{accepted only by approximate}}
\right)
=
\frac{\Dord}{2}.
\label{eq:dord_half}
\end{equation}
Thus $\Dord$ is the total probability of experimental outcomes for which the
two procedures make different confidence decisions about the tested value
$\mu$.

\subsection{The first-order phase structure}

At small $\lambda$ the calibrated ordering distance has the form
\begin{equation}
\boxed{
\Dord=a_1(\alpha,p)\lambda+o(\lambda).
}
\label{eq:a1_master}
\end{equation}
The coefficient $a_1$ can be understood directly as the probability moved
across the two acceptance boundaries by their first-order displacement.
Consider first the model-centered transformed Gaussian. 
Below, we use ``mod'' and ``dat'' to denote model-centered and
data-centered covariance reconstruction, respectively.  The usual
data-centered PPP construction corresponds to the untransformed member
\(p=1\) of this family. Expanding its
likelihood-ratio statistic around the exact one gives
\begin{equation}
q_{{\rm mod},p}-q_E
=
(p-1)\lambda Z
\left[
Z^2+(2+\alpha)H-4
\right]
+O(\lambda^2).
\label{eq:dq_mod}
\end{equation}
The detailed likelihood expansion is given in \cref{app:lr}.  What matters
for the ordering distance is its value at the Gaussian-limit acceptance
boundaries, $Z=\pm1$.  There the bracket reduces to
$-3+(2+\alpha)H$, and the two boundaries are displaced by the same amount,
\begin{equation}
Z_A^\pm-Z_E^\pm
=
-\frac{p-1}{2}
\left[-3+(2+\alpha)H\right]\lambda
+O(\lambda^2).
\label{eq:boundary_mod}
\end{equation}
Thus, at fixed ancillary disagreement $H$, the approximation translates the
acceptance interval sideways.  The probability exchanged at its two edges is
the boundary displacement times the standard-normal density $\phi(1)$,
summed over both edges.  Averaging over $H\sim\chi_1^2$ therefore gives
\begin{equation}
\boxed{
a_1^{\rm mod}(\alpha,p)
=
|p-1|\phione\,
\E\left|-3+(2+\alpha)H\right|.
}
\label{eq:a1_mod}
\end{equation}
For the data-centered construction the same boundary-shift calculation must
also include the first-order change of metric caused by evaluating the
covariance from the observed data.  The corresponding result is
\begin{equation}
a_1^{\rm dat}(\alpha,p)
=
2\phione\,
\E\left|
\frac{p-1}{2}
+
\left[
1-\alpha+\frac{p-1}{2}(2+\alpha)
\right]H
\right|,
\qquad
H\sim\chi_1^2.
\label{eq:a1_dat}
\end{equation}

Two limiting cases are particularly transparent:
\begin{align}
a_1^{{\rm dat},1}
&=2\phione(1-\alpha),
\label{eq:a1_ppp}
\\
a_1^{{\rm dat},\log}
&=\phione(1+3\alpha).
\label{eq:a1_log}
\end{align}
For PPP we have $p=1$, and the leading ordering discrepancy vanishes when the multiplicative component disappears, since \(1-\alpha=\beta\) is the normalization-uncertainty fraction.
The data-centered logarithm ($p=0$) does the opposite: its ordering distortion grows as the statistical fraction increases. It is also worth noting here that the square-root case develops a cancellation valley near $\alpha\simeq0.316$.

\begin{figure}[t]
\centering
\includegraphics[width=0.82\linewidth]{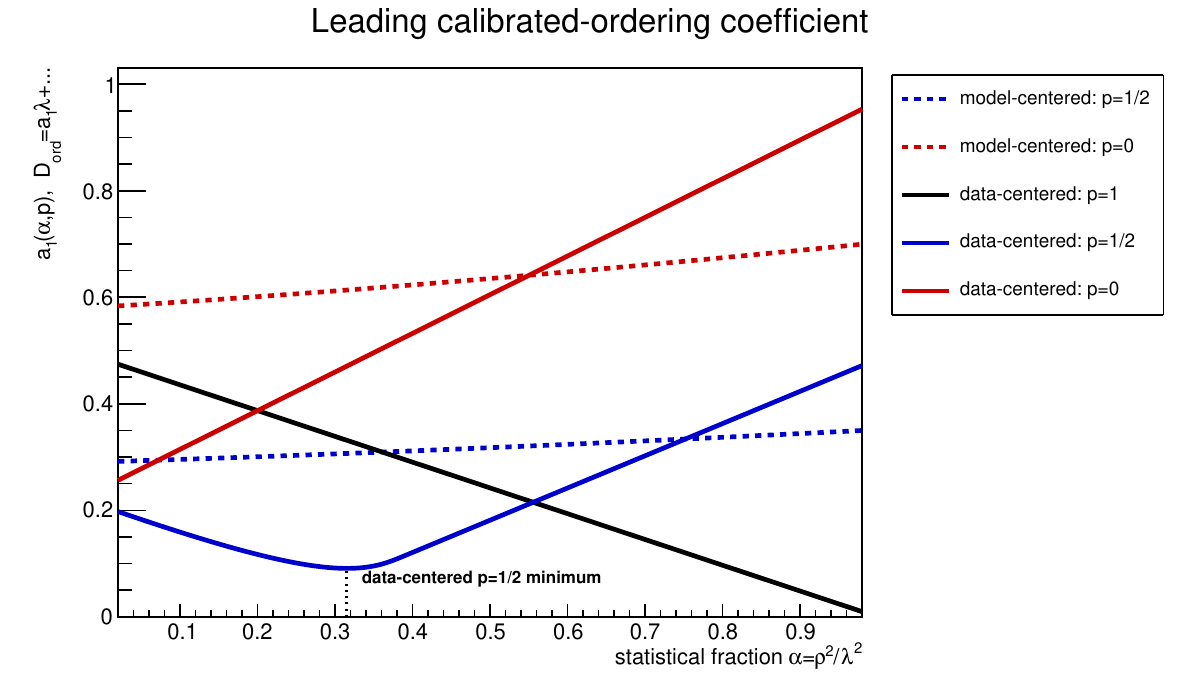}
\caption{Leading coefficient in
$\Dord=a_1(\alpha,p)\lambda+\cdots$.  Factoring out the overall uncertainty
scale exposes the mechanism: PPP ($p=1$) weakens toward the statistics-only
limit, the data-centered logarithm strengthens, and the square-root
prescription exhibits a cancellation valley.}
\label{fig:a1}
\end{figure}

\subsection{Exact coverage can hide ordering distortions}
\label{sec:coverage}

We ask here whether exact frequentist coverage is sufficient to show that an approximation has preserved the intended ordering of possible experimental outcomes. Focusing for definiteness on the likelihood-ratio ordering used throughout this paper\footnote{While the quantitative results discussed below are specific to likelihood-ratio ordering, the qualitative point that exact coverage alone does not ensure preservation of a chosen ordering prescription is general.}, we show that it is not: the covariance approximation can rearrange which data sets are accepted while leaving the total accepted probability unchanged.

Coverage and ordering answer different questions: coverage asks how much
probability the acceptance region contains; while an ordering prescription specifies which specific data sets make up that probability.

For the likelihood-ratio expansions considered here, let
\begin{equation}
\Dfix\equiv
\Pp_\mu(q_A\le 1)-\Pp_\mu(q_E\le 1)
\label{eq:Dfix_def}
\end{equation}
denote the difference in coverage obtained from the conventional fixed
threshold $q\le1$.  
This is distinct from the unconditional coverage of the
exactly calibrated belts, which is \(\gamma\) for both by construction.
The calibrated critical value has the form
\begin{equation}
c_\gamma=1+g_2\lambda^2+\cdots.
\end{equation}
The absence of a first-order correction has a simple geometric origin.
At \(O(\lambda)\) the deformation primarily translates the acceptance
interval: probability lost at one boundary is gained at the other, so
the total accepted probability is unchanged to first order.  The
critical value therefore needs to be recalibrated only at
\(O(\lambda^2)\), even though the identity of the accepted data sets
changes already at \(O(\lambda)\).

By contrast, the acceptance boundaries are translated already at first
order.  Consequently,
\begin{equation}
\boxed{
\Dord=O(\lambda),
\qquad
\Dfix=O(\lambda^2).
}
\label{eq:order_cov_hierarchy}
\end{equation}
The derivation and explicit \(g_2\) coefficients are given in
\cref{app:coverage}.

This hierarchy is not merely asymptotic bookkeeping.  It says that a coverage test can look excellent while a non-negligible fraction of the sample space is already being ordered differently and hence classified differently by the confidence construction.

\subsection{Coverage-blind points}
\label{sec:blind}

The contrast is sharpest when the $O(\lambda^2)$ coverage difference itself
vanishes.  For the data-centered logarithm,
\begin{equation}
\frac{\Dfix}{\phione\lambda^2}
=
\frac{25+6\alpha-57\alpha^2}{12},
\label{eq:log_covcoef}
\end{equation}
so the physical root is
\begin{equation}
\boxed{
\alpha_{\rm blind}^{{\rm dat},\log}
=
\frac{3+\sqrt{1434}}{57}
=
0.716986\ldots.
}
\label{eq:log_blind}
\end{equation}
At exactly the same point,
\begin{equation}
a_1^{{\rm dat},\log}
=
\phione(1+3\alpha_{\rm blind}^{{\rm dat},\log})
\simeq0.7624.
\end{equation}
Thus at $\lambda=0.1$,
\begin{equation}
\Dord\simeq7.6\%,
\end{equation}
while the leading conventional coverage difference is zero. We call this a coverage-blind point, because by only examining coverage one would be unable
to appraise the existence of a mismatch between which data is accepted by one
method and the other.

\begin{figure}[t]
\centering
\includegraphics[width=0.95\linewidth]{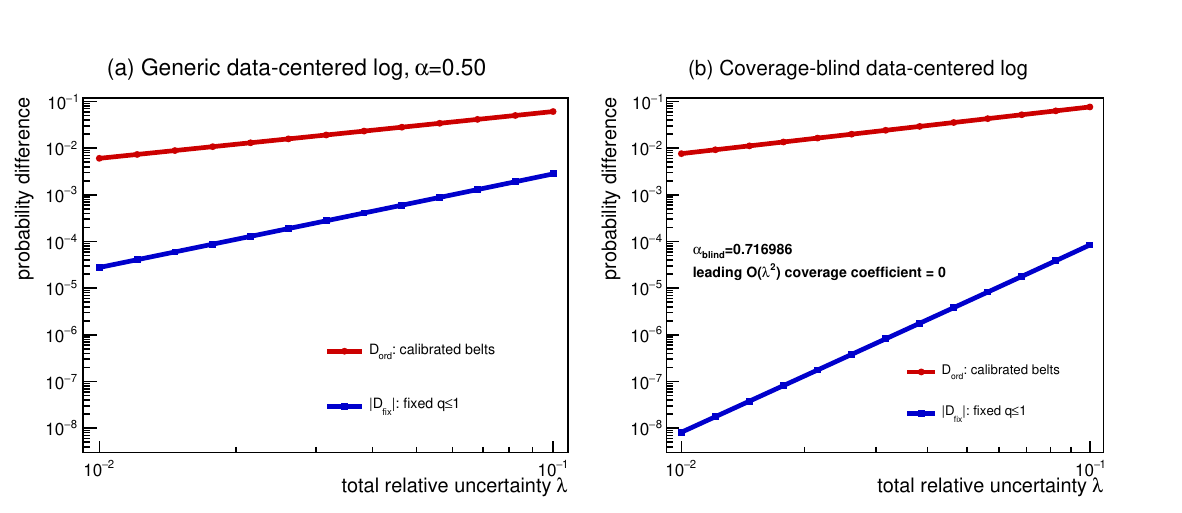}
\caption{
Scaling of ordering and fixed-threshold coverage discrepancies with the total relative uncertainty \(\lambda\) for the data-centered logarithmic reconstruction.
(a) At a generic point, \(\alpha=0.50\), the ordering distance \(\Dord\) follows the expected first-order behavior, \(\Dord\propto\lambda\), whereas the fixed-threshold coverage difference satisfies \(|\Dfix|\propto\lambda^2\).
(b) At the data-centered-log coverage-blind point, \(\alpha_{\rm blind}\equiv\alpha_{\rm blind}^{{\rm dat},\log}=0.716986\), the leading \(O(\lambda^2)\) coefficient of \(\Dfix\) vanishes, strongly suppressing the conventional coverage discrepancy, while \(\Dord\) remains first order in \(\lambda\). Thus agreement of a coverage diagnostic can coexist with a non-negligible change in likelihood-ratio ordering.
}
\label{fig:scaling}
\end{figure}

\section{Ancillarity, recognizable subsets, and what coverage averages away}
\label{sec:relevant}

The factorization in \cref{eq:factorization_pdf} makes $H$ a natural
recognizable ancillary subset variable.  This places the problem directly in
the tradition developed by Fisher, Cox, Buehler, Wallace, Kiefer, Robinson,
Casella, Goutis and others
\cite{Fisher1956,Cox1958,Buehler1959,Wallace1959,Kiefer1977,
Robinson1979,GoutisCasella1995}, and discussed in a high-energy-physics
setting by Cousins \cite{Cousins2011}.
Sundberg emphasized that ancillary conditioning improves relevance when
the ancillary acts as a precision index, and that not every ancillary has
this property \cite{Sundberg2003}.  In the exact model considered here,
$H$ does not index the conditional precision of $m$, since $m$ is
independent of $H$; the $H$-dependence of the confidence behavior is
introduced by the covariance approximation.

At fixed $H$, denote the exact and approximate acceptance regions by
$A_E(\mu)$ and $A_A(\mu)$, and define
\begin{align}
G(H)&=\Pp_\mu(A_A\setminus A_E\mid H),
\\
L(H)&=\Pp_\mu(A_E\setminus A_A\mid H).
\end{align}
Because the exact belt is independent of $H$,
$\Pp_\mu(A_E\mid H)=\gamma$.  Writing
\begin{equation}
C(H)\equiv\Pp_\mu(A_A\mid H),
\end{equation}
the two diagnostics are simply
\begin{equation}
\boxed{
\dcond(H)\equiv C(H)-\gamma=G(H)-L(H),
\qquad
\Dord(H)=G(H)+L(H).
}
\label{eq:gainloss}
\end{equation}

This identity is useful because it explains the perturbative hierarchy
geometrically.  A first-order sideways translation of an interval loses
$O(\lambda)$ probability on one edge and gains $O(\lambda)$ on the other.
Those changes add in $\Dord$ but cancel to first order in the coverage.

At fixed \(H\), the approximate and exact acceptance boundaries differ by an amount that can be expanded in powers of the small uncertainty scale \(\lambda\). We denote the corresponding first- and second-order displacement coefficients by \(d_1(H)\) and \(d_2(H)\). Then we may define the bookkeeping combination
\begin{equation}
e(H)\equiv d_2(H)-\beta d_1(H)-\frac12 d_1(H)^2.
\end{equation}
The quantity \(e(H)\) is not a new statistic; it simply collects the
second-order boundary terms that survive in the conditional coverage.
Away from zeros of the first-order boundary displacement $d_1(H)$, the local
formulas are
\begin{align}
\dcond(H)
&=
2\phione\lambda^2 e(H)+O(\lambda^4),
\label{eq:condcov}
\\
\Dord(H)
&=
2\phione\lambda |d_1(H)|+O(\lambda^3).
\label{eq:condord}
\end{align}
For the PPP construction,
\begin{equation}
d_1(H)=\beta H,
\end{equation}
so the ordering mismatch grows with the disagreement even though $H$ carries
no information about $\mu$.
The endpoint \(H=0\) is exceptional because there the first-order
displacement itself vanishes.  The resulting nonuniform boundary layer
and its \(O(\lambda^{5/2})\) contribution to the unconditional ordering
distance are discussed in \cref{app:lr}.

\begin{figure}[t]
\centering
\begin{subfigure}[t]{0.49\linewidth}
\includegraphics[page=1,width=\linewidth]{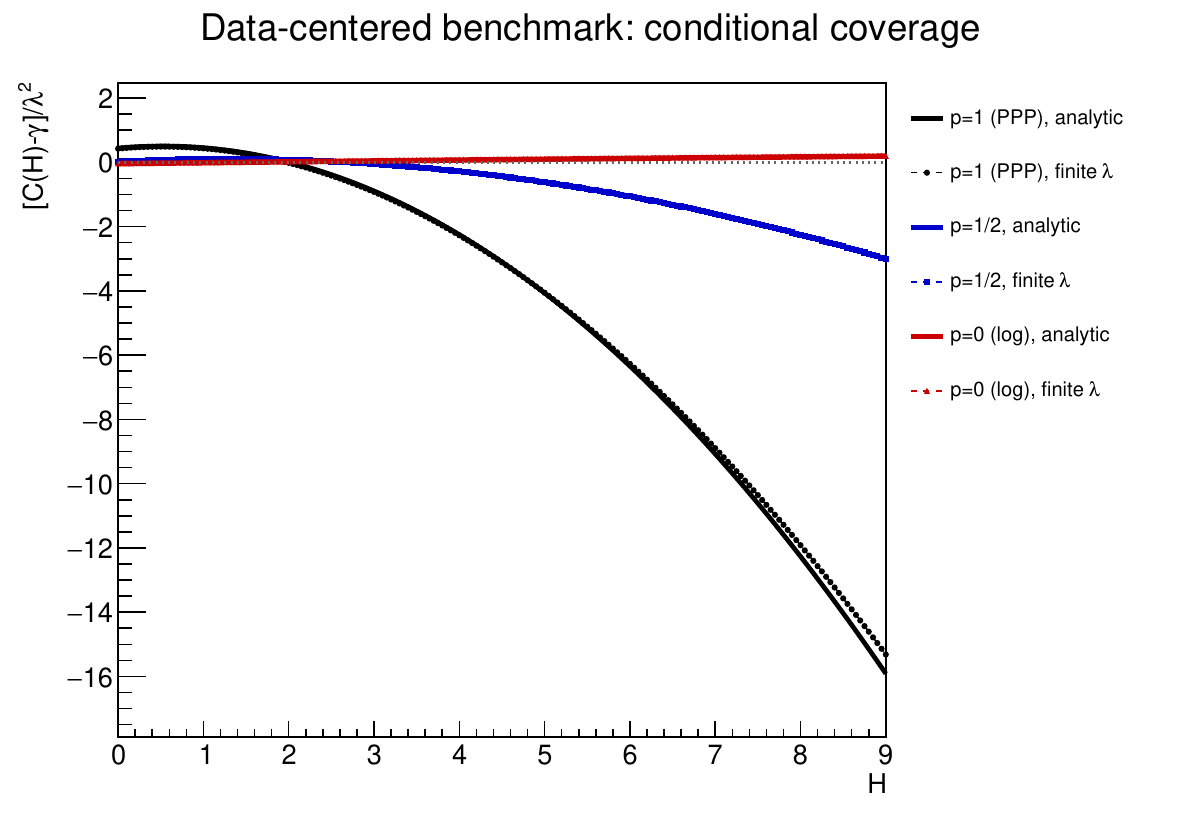}
\caption{Conditional coverage defect.}
\end{subfigure}\hfill
\begin{subfigure}[t]{0.49\linewidth}
\includegraphics[page=2,width=\linewidth]{fig_relevant.pdf}
\caption{Conditional ordering mismatch.}
\end{subfigure}
\caption{Conditioning on the ancillary residual disagreement.  Coverage is a
signed imbalance and can remain small or change sign; the ordering distance
measures the total mass reassigned and responds one perturbative order
earlier.}
\label{fig:conditional}
\end{figure}

A globally calibrated procedure can therefore hide structure in two stages.
First, gained and lost probability cancel at fixed $H$.  Second, the approximation may overcover for some ranges of the ancillary disagreement \(H\) and undercover for others, with the two effects cancelling when coverage is averaged over all experiments:
\begin{equation}
\E_H[\dcond(H)]=0.
\end{equation}

\subsection{Recognizable and relevant subsets}

Subsets such as
\begin{equation}
\mathcal S(h_0)=\{H>h_0\}
\end{equation}
are recognizable after observing the experiment.  They show directly how the
conditional coverage and ordering change as one selects increasingly
discrepant measurements.

The formal Buehler notion of a negatively biased relevant subset is stronger: one fixed subset of the observation space must generate confidence intervals whose conditional coverage remains below the nominal level by a finite amount, for every value of the parameter\cite{Buehler1959,Cousins2011}.  
In other words, if we restrict attention to experiments whose data fall in that same recognizable subset, the resulting confidence intervals must contain the true parameter too rarely, whatever its true value.

The simplest monotone high-$H$ or low-$H$ subsets do not satisfy that requirement for PPP across the entire physical regime.  A short leading-order reconnaissance nevertheless
shows that disconnected ancillary subsets combining extremely small and very
large $H$ can satisfy the perturbative conditions for uniform negative
conditional bias.  Because such a subset is mathematically legitimate but
less natural as a conditioning argument, we regard it as supporting evidence
rather than the central physical message we wish to pass.
Closed-form results for monotone ancillary tail subsets are given in
\cref{app:tails}. A leading-order analysis of the stronger Buehler relevant-subset condition,
together with a finite-$s$ numerical example for a fixed disconnected
ancillary subset, is given in \cref{app:buehler}. 
\section{What changes in the confidence interval one actually reports?}
\label{sec:intervals}

The ordering language becomes experimentally concrete when the acceptance regions are turned into confidence intervals. For each candidate value of \(\mu\), the Neyman construction defines the set of possible data that would be accepted. Once the data \(x\) are observed, we reverse this question: we retain all values of \(\mu\) whose acceptance region contains \(x\). This inversion gives
$$ I_E(x)=\{\mu:x\in A_E(\mu)\}, \qquad I_A(x)=\{\mu:x\in A_A(\mu)\}. $$
Thus two equally calibrated belts can yield different confidence intervals for exactly the same observed data.

The cleanest diagnostic for ordering issues is to keep the informative statistic $m$ fixed and change only the ancillary $H$.  The exact marginal likelihood then remains unchanged, so the exact confidence interval must be identical.  Any movement of an approximate interval is direct ancillary leakage.

\subsection{A coverage-blind example}

To make a concrete example of coverage blindness, let us consider the logarithmic representation
\begin{equation}
h(x)=\log x,
\end{equation}
using the data-centered covariance prescription, in which the Jacobian
$h'(x_i)=1/x_i$ is evaluated at the observed central values.  At its
coverage-blind point,
\begin{equation}
\alpha=0.716986,\qquad \lambda=0.1,\qquad \mu_0=1.
\end{equation}
For three data sets with the same $m=\mu_0$ but different disagreement,
the exact $68.27\%$ interval is always
\begin{equation}
I_E=[0.90015,\,1.09987].
\end{equation}
The approximate log intervals are
\begin{align}
H=0:\quad &[0.90480,\,1.10500],
\\
H=1:\quad &[0.91476,\,1.11656],
\\
H=4:\quad &[0.94438,\,1.15093].
\end{align}
At $H=4$ the approximate interval center has shifted upward by about
$4.8\%$ of $\mu_0$, while its length changes much less.

\begin{figure}[t]
\centering
\includegraphics[page=4,width=0.88\linewidth]{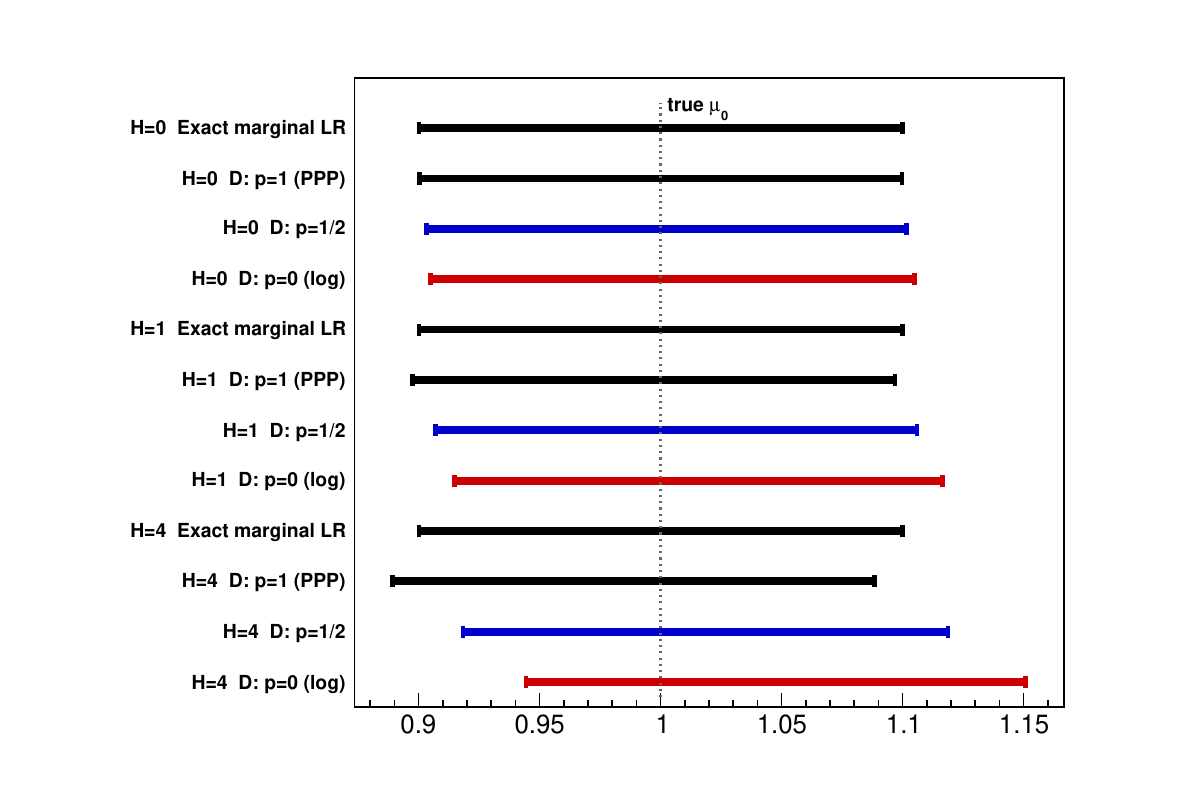}
\caption{Same informative weighted mean, different ancillary disagreement. With increasing $H$, the exact interval is
unchanged, while the covariance-based log interval moves, even though both confidence constructions are calibrated to the
same nominal coverage. In the legend, ``D'' denotes the data-centered reconstruction.}
\label{fig:intervals}
\end{figure}

A $10^6$-toy inversion provides an independent check of the operational
meaning of $\Dord$.  At the blind point the data-centered log confidence interval
changes its decision about the true $\mu_0$ in
\begin{equation}
7.652\%\quad\text{of experiments},
\end{equation}
in agreement with the direct acceptance-region calculation.
For comparison, the corresponding swap probabilities are approximately
$1.37\%$ for PPP and $3.22\%$ for the square-root transformation.

The two-measurement PPP benchmark introduced in \cref{s:3descr}
is even more striking.  With
$m=10.5$ fixed, the exact interval remains
\begin{equation}
[8.420,\,12.601],
\end{equation}
whereas the $p=1$ PPP interval changes from approximately
$[8.342,12.662]$ at $H=0$ to $[7.045,11.063]$ at $H=4$.
The dominant effect is a translation of the interval rather than a large
change of its width.

The practical point emerging from the above discussion is not that equal coverage permits different orderings—it plainly does—but that the covariance approximation changes the ordering produced by the same likelihood-ratio principle. Even after the exact and approximate constructions are separately calibrated to the same coverage, this change leads to different confidence intervals for a non-negligible fraction of experiments.

\section{Beyond two measurements: the arbitrary-$N$ structure}
\label{sec:robustness}

The two-measurement model discussed above has been illustrative because its geometry can be drawn: one direction carries the weighted mean, while the orthogonal direction measures the disagreement of the two inputs.  It is natural, however, to ask whether the results we derived rely on this particularly simple geometry.  In a real combination with $N$ measurements there are $N-1$ independent residual directions, and there is no a priori reason for a covariance approximation to treat all of them through a single ancillary quantity. However, for the $p=1$ data-centered PPP prescription, remarkably, it does:
the entire problem continues to reduce to the weighted mean and the usual
goodness-of-fit chi-square.  This allows the ordering and coverage results to
be extended analytically to arbitrary finite $N$ and arbitrary statistical
variances.

\subsection{The informative direction and the ancillary residual space}

Let us define
\begin{equation}
\Vstat=\diag(\sigma_1^2,\ldots,\sigma_N^2),
\qquad
W=\Vstat^{-1},
\end{equation}
 and 
\begin{equation}
\omega=\mathbf1^\trans W\mathbf1,
\qquad
\tau^2=\omega^{-1},
\qquad
m=\tau^2\mathbf1^\trans Wx.
\end{equation}
where \(x=(x_1,\ldots,x_N)^T\) is the vector of observed measurements and \(\mathbf1\) is the \(N\)-component vector of ones.
Writing
\begin{equation}
\mathbf r=x-m\mathbf1,
\qquad
H=\mathbf r^\trans W\mathbf r,
\end{equation}
one has by construction
\begin{equation}
\mathbf1^\trans W\mathbf r=0.
\end{equation}

The meaning of these quantities is the same as in the two-measurement
example, but the geometry has changed.  The scalar $m$ specifies the position
of the data along the common-mean direction, whereas $\mathbf r$ lives in an
$(N-1)$-dimensional residual subspace and describes all possible internal
disagreements among the measurements.

For the exact Gaussian model,
\begin{equation}
m\mid\mu\sim
\mathcal N\!\left(\mu,\tau^2+s^2\mu^2\right),
\qquad
m\perp\mathbf r,
\end{equation}
while
\begin{equation}
H\sim\chi^2_{\nu},
\qquad
\nu=N-1.
\label{eq:H_general_chi2}
\end{equation}
The shared multiplicative fluctuation therefore affects only the informative
direction $m$.  The complete residual vector is ancillary for $\mu$.
This is already enough to show that the separation found for two
measurements was not a consequence of having only one residual direction.
What matters for the extension below is an additional simplification: the
data-centered covariance approximation also sees the residual space only
through its squared length $H$.

\subsection{The PPP likelihood for arbitrary $N$}

Consider again the data-centered covariance
\begin{equation}
\widehat V_x=\Vstat+s^2xx^\trans.
\label{eq:general_plugin_cov}
\end{equation}
Using the Sherman--Morrison identity and the decomposition \(x=m\mathbf1+\mathbf r\), one obtains the exact reduction
\begin{equation}
\boxed{
Q_{\rm PPP}(\mu)
=
\frac{H}{1+s^2H}
+
\frac{\omega(1+s^2H)}
     {1+s^2H+\omega s^2m^2}
\left[
\mu-\frac{m}{1+s^2H}
\right]^2 .
}
\label{eq:general_plugin_Q}
\end{equation}
with the algebra given in \cref{app:estimators}.

\Cref{eq:general_plugin_Q} is the key to the $N$-measurement extension.  Although the ancillary residual space has dimension $N-1$, neither its orientation nor its
individual components enter the likelihood-ratio construction: all of its
influence is compressed into the single statistic $H$.

Minimizing \cref{eq:general_plugin_Q} therefore gives
\begin{equation}
\muhat_{{\rm dat},1}
=
\frac{m}{1+s^2H},
\label{eq:general_plugin}
\end{equation}
exactly as for two measurements.  The arbitrary-$N$ point-estimate structure
is consistent with the analysis of Neudecker, Fr\"uhwirth and Leeb
\cite{NeudeckerFruehwirthLeeb2012,NeudeckerEtAl2014}; the extension developed here concerns its
ancillary interpretation and its consequences for likelihood-ratio ordering
and coverage.  A larger internal goodness-of-fit statistic produces a larger
displacement of the fitted common mean even though $H$ is ancillary in the
exact model.

The same reduction also makes it possible to go beyond the point estimate
and derive the likelihood-ratio ordering itself, as we now show.

\subsection{Ordering for arbitrary $N$}

We use the same local variables introduced in
\cref{sec:ordering}, now with $\tau$ denoting the statistical uncertainty of
the weighted mean for the full $N$-measurement combination.  After scaling by
the tested true value $\mu_0$,
\begin{equation}
m=1+\lambda Z,
\qquad
Z\sim\mathcal N(0,1),
\qquad
Z\perp H,
\end{equation}
with
\begin{equation}
\lambda^2=s^2+\rho^2,
\qquad
\rho=\frac{\tau}{\mu_0},
\qquad
\alpha=\frac{\rho^2}{\lambda^2},
\qquad
\beta=\frac{s^2}{\lambda^2}=1-\alpha.
\end{equation}

The exact likelihood still depends only on $m$, so its local
likelihood-ratio expansion is unchanged.  For the data-centered PPP
prescription, expansion of \cref{eq:general_plugin_Q} instead gives
\begin{equation}
\boxed{
q_{\rm PPP}-q_E
=
-2\beta\lambda Z
\left(Z^2+H-1\right)
+O(\lambda^2),
}
\label{eq:general_ppp_lr_difference}
\end{equation}
the same functional form obtained for two measurements, but now with
$H\sim\chi^2_{N-1}$.

The physical content of this result is particularly transparent at the
Gaussian-limit $68.27\%$ boundary.  At fixed $H$, the two edges of the PPP
acceptance interval are shifted relative to the exact ones by
\begin{equation}
\boxed{
\delta Z(H)
=
\beta H\lambda+O(\lambda^2).
}
\label{eq:general_boundary_shift}
\end{equation}
Thus an ancillary quantity that measures only the internal consistency of
the measurements translates the acceptance interval used to infer their
common mean.

The ordering distance at fixed $H$ is consequently
\begin{equation}
\Dord(H)
=
2\phi(1)\,\beta H\,\lambda+o(\lambda),
\end{equation}
and averaging over the ancillary distribution,
$\E[H]=N-1$, gives
\begin{equation}
\boxed{
\Dord
=
2\phi(1)\,\beta\,(N-1)\lambda+o(\lambda).
}
\label{eq:general_ppp_dord}
\end{equation}

For $N=2$, \cref{eq:general_ppp_dord} reduces to
$a_1=2\phi(1)(1-\alpha)$, as obtained in
\cref{eq:a1_ppp}.  The factor $N-1$ has a simple interpretation: it is the
mean ancillary goodness-of-fit, $\E[H]$, associated with the $N-1$ residual
degrees of freedom.

The individual statistical uncertainties \(\sigma_i\) therefore affect the leading ordering result only through the total statistical precision \(\tau\). At fixed \(\tau\), changing their relative sizes changes the geometry of the residual space but not the distribution \(H\sim\chi^2_{N-1}\). Thus the leading result does not require equal statistical errors and applies to arbitrary heteroscedastic \(\sigma_i\).

\subsection{Coverage blindness without a nonlinear transformation}

The arbitrary-$N$ extension also reveals a feature that is absent in the
two-measurement $p=1$ PPP example.  Carrying the fixed-threshold
coverage calculation to second order gives, with $\nu=N-1$,
\begin{equation}
\boxed{
\frac{\Dfix}{\phi(1)\lambda^2}
=
\beta
\left[
2\nu+3\beta-\beta\nu(\nu+3)
\right]
+O(\lambda^2).
}
\label{eq:general_ppp_coverage}
\end{equation}
For $\nu=1$ this reduces to
\begin{equation}
\frac{\Dfix}{\phi(1)\lambda^2}
=
\beta(2-\beta)
=
1-\alpha^2,
\end{equation}
recovering the two-measurement result of
\cref{app:coverage}.

For $N=2$ the only nontrivial zero of the leading coefficient would require
$\beta=2$ and is therefore outside the physical range.  Starting from three
measurements, however, a physical coverage-blind solution appears:
\begin{equation}
\boxed{
\beta_{\rm blind}(N)
=
\frac{2\nu}
     {\nu(\nu+3)-3},
\qquad
\nu=N-1.
}
\label{eq:general_ppp_blind}
\end{equation}
For $N=3$ this gives
\begin{equation}
\beta_{\rm blind}=\frac47,
\qquad
\alpha_{\rm blind}=\frac37.
\end{equation}
At the same point the ordering coefficient remains
\begin{equation}
a_1
=
2\phi(1)\beta_{\rm blind}(N-1)
=
\frac{16}{7}\phi(1)
\simeq0.5531 .
\end{equation}
Thus, for $\lambda=0.1$, the leading prediction is
\begin{equation}
\Dord\simeq5.53\%,
\end{equation}
even though the $O(\lambda^2)$ fixed-threshold coverage difference vanishes.

This shows that the coverage-blind phenomenon of \cref{sec:blind} is not tied to nonlinear representation changes. Nonlinear representation changes provide a particularly clear two-measurement realization of the phenomenon, but they are not required for it: with three or more inputs, the $p=1$ PPP construction already exhibits the same separation between coverage and ordering.

The exact finite-$\lambda$ calculation confirms that this is not merely an
algebraic cancellation.  For $N=3$, $\lambda=0.1$, and
$\beta=4/7$, we find
\begin{equation}
\Dord=0.0539,
\end{equation}
while the fixed-threshold coverage difference is only
$1.7\times10^{-5}$.  The corresponding $N=4$ blind point,
$\beta=0.4$, gives $\Dord=0.0568$.  Details of the numerical validation are
given in \cref{app:numerics}.

\subsection{Unequal errors and nonlinear representations}

The simplification above is exact for the untransformed ($p=1)$ PPP prescription.
For nonlinear representations the situation is richer, because a
componentwise transformation can distinguish different directions within the
ancillary residual space.

\begin{figure}[t]
\centering
\begin{subfigure}[t]{0.49\linewidth}
\includegraphics[page=1,width=\linewidth]{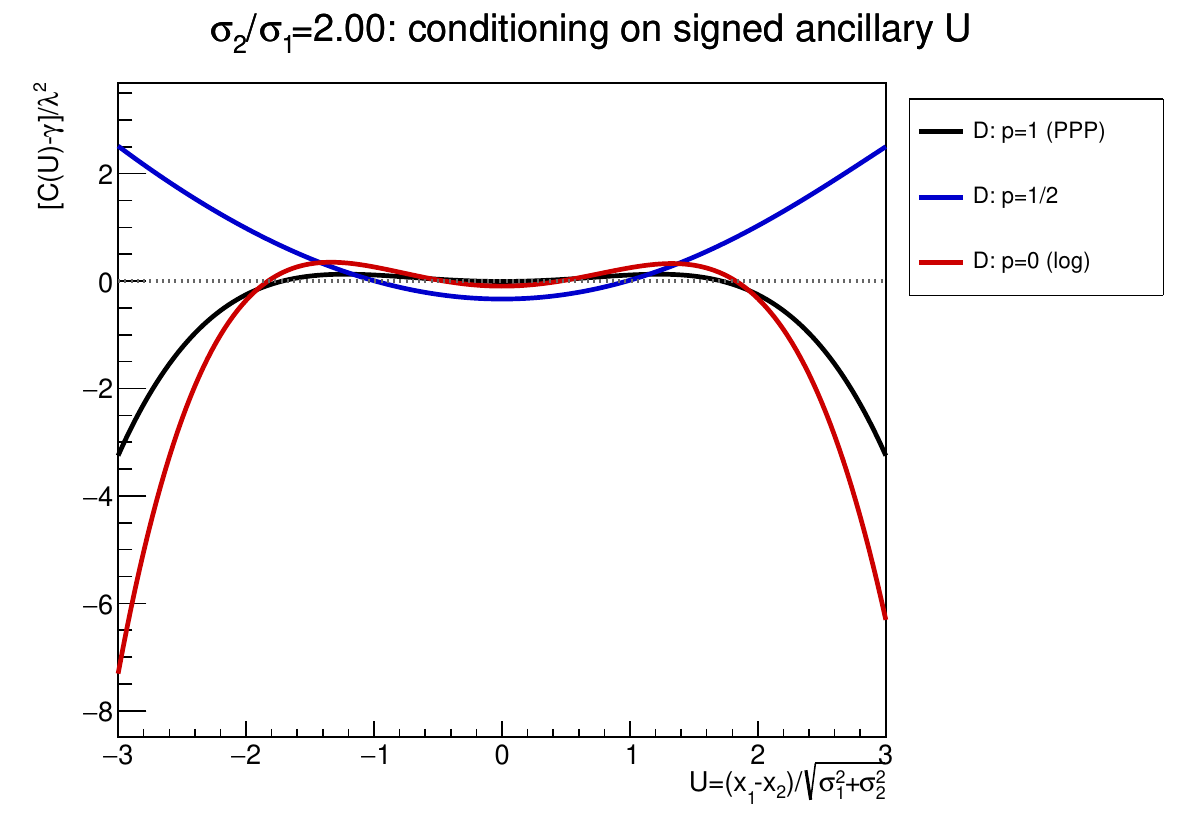}
\caption{Conditional coverage versus signed $U$.}
\end{subfigure}\hfill
\begin{subfigure}[t]{0.49\linewidth}
\includegraphics[page=2,width=\linewidth]{FC_unequal_r2_ancillary.pdf}
\caption{Ordering mismatch versus signed $U$.}
\end{subfigure}

\caption{Example with $\sigma_2/\sigma_1=2$.  The PPP construction retains the
$U\leftrightarrow-U$ symmetry, whereas nonlinear covariance reconstructions
acquire a directional ancillary dependence.  The unconditional leading
ordering effect is nevertheless insensitive to the variance ratio at fixed
total precision. In the legends, ``D'' denotes the data-centered reconstruction.}
\label{fig:unequal}
\end{figure}

Already for two measurements, unequal statistical errors break the exchange
symmetry of the signed ancillary
\begin{equation}
U=
\frac{x_1-x_2}{\sqrt{\sigma_1^2+\sigma_2^2}}.
\end{equation}
Holding the total statistical precision $\tau$ fixed while varying
$r_\sigma=\sigma_2/\sigma_1$ leaves the $p=1$ PPP problem exactly unchanged.
For the nonlinear prescriptions considered above, the leading unconditional
ordering coefficient is also independent of $r_\sigma$, while unequal errors
enter at higher order.  They can nevertheless make the inference distinguish
$+U$ from $-U$, although both signs remain ancillary for $\mu$.  For the signed-ancillary display in \cref{fig:unequal}, we write
\(C(U)\equiv \Pp_\mu(A_A\mid U)\).  The figure makes the directional
effect explicit: for the nonlinear reconstructions the conditional
coverage and ordering can differ between \(+U\) and \(-U\), whereas the
\(p=1\) PPP construction retains the \(U\leftrightarrow -U\) symmetry.

The arbitrary-$N$ result and the unequal-error example therefore establish
two complementary forms of robustness.  For the untransformed PPP
construction, the ordering and coverage effects can be derived analytically
for arbitrary finite $N$ and arbitrary $\sigma_i$.  Nonlinear
representations need not preserve the radial symmetry of the ancillary
subspace, but the two-measurement unequal-error example shows that the
ordering distortion itself is not generated by the exchange symmetry of the
simplest model.

\section{Discussion and implications for experimental practice}
\label{sec:practice}

The results discussed above suggest a progressively more stringent set of
questions for approximate likelihoods.

\paragraph{Does the approximation distort the point estimate?}
A natural first diagnostic is whether the approximate construction produces
a strongly displaced or biased estimator, or degrades its mean-squared error.
This is useful but insufficient.  The exact marginal MLE in our model
need not itself be unbiased, so estimator bias alone does not diagnose
model misspecification.

\paragraph{Does the procedure cover?}
Coverage is essential, but it fixes only the probability content of an acceptance region. Exact calibration does not establish that the exact and approximate constructions make the same accept/reject decisions for individual experimental outcomes. At a coverage-blind point, even an apparently
perfect coverage comparison can miss a first-order ordering change.

\paragraph{Is the inference stable under physically meaningful
reparameterizations?}
An exact likelihood has this invariance: a one-to-one change of variable cannot alter likelihood-ratio inference. Instead, a reconstructed Gaussian need not share this property.  Rate-to-coupling and lifetime-to-width transformations therefore provide useful stress tests when only central values and covariance
information are available.

\paragraph{Does an ancillary statistic enter parameter inference?}
When a natural ancillary exists, conditioning provides a particularly sharp
diagnostic.  In our model $H$ measures only the internal disagreement of the
measurements.  The exact likelihood does not use it for inference on $\mu$;
the covariance approximation can.  Dependence of the inferred parameter on
such a quantity is direct evidence that the approximation has introduced
information that was absent from the exact parameter likelihood.

\paragraph{Can the ordering itself be compared?}
When an exact likelihood, or another well-justified reference construction,
is available, $\Dord$ provides a direct measure of how much the approximate
ordering differs from it: it is the probability, under the reference model,
of experimental outcomes for which the two constructions make different
confidence decisions. 

The strongest practical remedy is to preserve and publish the statistical
model rather than only central values and a covariance matrix.  
This message is not new; the value of publishing full or simplified
likelihood information, and of treating simplified likelihoods explicitly
as controlled approximations, is well established in particle physics
\cite{BuckleyEtAl2019,Kieseler2017,CranmerEtAl2022,Berger2023}.
Here we have offered a characterization of what can be lost when such a model is unavailable. We hope we have also clarified the role of Feldman--Cousins ordering in this context. The ordering prescription is not the source of the problem; indeed, applying it to the exact likelihood we retain the ancillary factorization.  
Applying it to an approximate likelihood, the method also faithfully
uses the likelihood-ratio ordering implied by that approximation; any
ancillary dependence or representation dependence introduced by the
approximation is therefore propagated into the resulting acceptance regions.

\section{Conclusions}
\label{sec:conclusion}
A covariance matrix can be a faithful summary of a linear Gaussian measurement, but it is not in general a representation-invariant substitute for a full statistical model.

In the correlated-measurement problem studied here, the data decompose exactly into an informative weighted mean and an ancillary residual disagreement. Data-dependent covariance matrices and nonlinear Gaussian reconstructions can break that separation. The result is not merely a displaced estimator: the likelihood-ratio ordering of possible experimental outcomes changes.

The probability mass of the symmetric difference of equally calibrated acceptance regions,
\begin{equation}
\Dord=\Pp_\mu(A_A\triangle A_E),
\end{equation}
has a direct operational meaning as the fraction of experiments whose confidence decision changes. The ordering distance is first order in the total relative uncertainty $\lambda$ (combining the statistical and normalization components), $\Dord=O(\lambda)$, whereas both the fixed-threshold and ancillary-conditioned coverage defects begin only at second order, $O(\lambda^2)$. Consequently, for some covariance reconstructions there are uncertainty compositions at which the leading $O(\lambda^2)$ coverage coefficient vanishes while a nonzero first-order ordering discrepancy remains.

In the language of relevant-subset theory, the covariance approximation manufactures inferential relevance for a goodness-of-fit statistic that is ancillary in the true model. In practical terms, data sets with the same informative weighted mean $m$ but different ancillary disagreement $H$ can then be ordered differently; after inversion, this can lead to different confidence intervals.

The practical conclusion is therefore narrower, but stronger, than a generic warning about covariance matrices: coverage tests alone cannot certify that an approximate covariance likelihood preserves the intended likelihood-ratio ordering of the underlying measurement model.

\section*{Acknowledgments}

The author acknowledges the assistance from OpenAI's ChatGPT (GPT-5.6 Sol) in the development and checking of analytic derivations, numerical studies and code, literature analysis, and preparation of the
manuscript. The scientific direction, interpretation of the results, and final responsibility for the content remain with the author.
\clearpage

\begin{appendices}
\crefalias{section}{appendix}

\section{Exact marginal and data-centered estimators in the untransformed case}
\label{app:estimators}

For the model in \cref{eq:model}, the exact marginal likelihood is
\begin{equation}
-2\log L_{\rm marg}
=
\frac{\omega(m-\mu)^2}{1+\omega s^2\mu^2}
+
\log(1+\omega s^2\mu^2)+\mathrm{const}.
\end{equation}
The stationary point of $L_{marg}$ above obeys
\begin{equation}
\tau^2m+
\left[s^2m^2-(1+s^2)\tau^2\right]\mu
-s^2m\mu^2-s^4\mu^3=0.
\end{equation}
The root corresponding to the MLE simplifies, in the systematic-dominated limit, to
\begin{equation}
\muhat_{\rm marg}
=
\frac{2m}{1+\sqrt{1+4s^2}}
=
m(1-s^2+2s^4+\cdots).
\end{equation}

\paragraph{Data-centered covariance for arbitrary $N$.}
Let us write
\begin{equation}
\widehat V_x=\Vstat+s^2xx^\trans.
\end{equation}
For an invertible matrix \(A\) and vectors \(u,v\), the Sherman--Morrison identity states
\begin{equation}
(A+uv^\trans)^{-1}
=
A^{-1}
-
\frac{A^{-1}uv^\trans A^{-1}}
     {1+v^\trans A^{-1}u},
\end{equation}
provided \(1+v^\trans A^{-1}u\neq0\). Applying it here with
\(A=\Vstat\) and \(u=v=sx\) gives
\begin{equation}
\widehat V_x^{-1}
=
W-\frac{s^2Wxx^\trans W}
        {1+s^2x^\trans Wx}.
\end{equation}
Writing
\begin{equation}
x=m\mathbf1+\mathbf r,
\qquad
\mathbf1^\trans W\mathbf r=0,
\qquad
H=\mathbf r^\trans W\mathbf r,
\end{equation}
one has
\begin{align}
x^\trans Wx
&=\omega m^2+H,\\
(x-\mu\mathbf1)^\trans Wx
&=\omega m(m-\mu)+H,\\
(x-\mu\mathbf1)^\trans W(x-\mu\mathbf1)
&=\omega(m-\mu)^2+H.
\end{align}
Substitution and completion of the square then give, for arbitrary $N$ and
arbitrary statistical variances,
\begin{equation}
\boxed{
Q_{\rm PPP}(\mu)
=
\frac{H}{1+s^2H}
+
\frac{\omega(1+s^2H)}
     {1+s^2H+\omega s^2m^2}
\left[
\mu-\frac{m}{1+s^2H}
\right]^2 .
}
\label{eq:app_general_plugin_Q}
\end{equation}
Its minimum is therefore
\begin{equation}
\muhat_{{\rm dat},1}
=
\frac{m}{1+s^2H}.
\end{equation}

For $N$ measurements,
\begin{equation}
H\sim\chi^2_\nu,
\qquad
\nu=N-1,
\end{equation}
and $m\perp H$.  Hence
\begin{equation}
\E[\muhat_{{\rm dat},1}]
=
\mu B_\nu(s),
\qquad
B_\nu(s)
\equiv
\E\!\left[\frac{1}{1+s^2H}\right].
\end{equation}
The expectation can be written as
\begin{equation}
B_\nu(s)
=
(2s^2)^{-\nu/2}
e^{1/(2s^2)}
\Gamma\!\left(
1-\frac{\nu}{2},
\frac{1}{2s^2}
\right),
\label{eq:Bnu}
\end{equation}
where $\Gamma(a,x)$ is the upper incomplete gamma function.  Its
small-$s$ expansion is
\begin{equation}
B_\nu(s)
=
1-\nu s^2+\nu(\nu+2)s^4+O(s^6).
\end{equation}
Thus the leading relative bias is
\begin{equation}
\frac{\E[\muhat_{{\rm dat},1}]-\mu}{\mu}
=
-(N-1)s^2+O(s^4).
\end{equation}

For $N=2$, $\nu=1$, \cref{eq:Bnu} reduces to
\begin{equation}
B_1(s)
=
\sqrt{\frac{\pi}{2s^2}}\,
e^{1/(2s^2)}
\operatorname{erfc}\!\left(\frac{1}{\sqrt{2}s}\right),
\end{equation}
which is the exact expectation obtained by averaging the two-measurement
untransformed data-centered estimator of \cref{eq:plugin_est} over $H\sim\chi^2_1$.

\section{Transformed Gaussian likelihoods}
\label{app:transform}

For a smooth componentwise transformation $h$, let $Y=h(X)$ componentwise
and define, for the observed data,
\begin{equation}
y_i=h(x_i),
\qquad
\bar y=\frac{\sum_i w_i y_i}{\omega},
\qquad
R_h=\sum_i w_i(y_i-\bar y)^2,
\qquad
v(\mu)=\tau^2+s^2\mu^2.
\end{equation}
The model-centered Gaussian reconstruction has
\begin{equation}
Y\mid\mu \,
\dot\sim \,
\mathcal N\!\left(
h(\mu)\mathbf1,\,
h'(\mu)^2[\Vstat+s^2\mu^2\mathbf1\mathbf1^\trans]
\right).
\end{equation}
Above, \(\dot\sim\) means ``is approximately distributed as'';
we use it to distinguish the locally reconstructed Gaussian distribution
from the exact sampling distribution.
For two measurements its normalized log likelihood can be written
\begin{equation}
-2\widetilde\ell_h
=
\frac{R_h}{h'(\mu)^2}
+
\frac{[\bar y-h(\mu)]^2}{h'(\mu)^2v(\mu)}
+
4\log|h'(\mu)|
+\log v(\mu)+\mathrm{const}.
\end{equation}

For Box--Cox $h_p$, let
\begin{equation}
u_i=x_i^p,
\qquad
\bar u=\frac{\sum_i w_i u_i}{\omega},
\qquad
Q_u=\sum_i w_i(u_i-\bar u)^2.
\end{equation}
Then
\begin{equation}
-2\widetilde\ell_p
=
\frac{Q_u+(\bar u-\mu^p)^2/v(\mu)}
     {p^2\mu^{2p-2}}
+
4(p-1)\log\mu+\log v(\mu)+\mathrm{const}.
\end{equation}

For the data-centered Jacobian, define the relative covariance
\begin{equation}
\widetilde V(x)
=
V_{\rm rel}(x)+s^2\mathbf1\mathbf1^\trans,
\qquad
[V_{\rm rel}(x)]_{ii}=\frac{\sigma_i^2}{x_i^2}.
\end{equation}
The coordinate dependence then appears in the residual map
\begin{equation}
\delta_{p,i}(\mu)=
\frac{1-(\mu/x_i)^p}{p},
\qquad
\delta_{0,i}(\mu)=\log(x_i/\mu).
\end{equation}
For the data-centered reconstruction the propagated covariance is fixed once
the data have been observed, so its determinant contributes only a
$\mu$-independent constant.  In the notation above,
\begin{equation}
-2\widetilde\ell^{\,\rm dat}_p(\mu)
=
\boldsymbol{\delta}_p(\mu)^\trans
\widetilde V(x)^{-1}
\boldsymbol{\delta}_p(\mu)
+\mathrm{const.},
\label{eq:app_dat_likelihood}
\end{equation}
where $\boldsymbol{\delta}_p$ denotes the vector with components
$\delta_{p,i}$.

\section{Likelihood-ratio expansions}
\label{app:lr}

The exact marginal likelihood ratio (LR) at the true value has
\begin{equation}
q_E
=
Z^2
-2\beta\lambda Z
+
\beta\left[(2\beta-1)Z^2+\beta\right]\lambda^2
+O(\lambda^3).
\label{eq:app_qE}
\end{equation}
For the model-centered Box--Cox reconstruction, writing
\begin{equation}
\kappa=p-1,
\end{equation}
the first difference from the exact LR is
\begin{equation}
\boxed{
q_{{\rm mod},p}-q_E
=
\kappa\lambda Z
\left[
Z^2+(2+\alpha)H-4
\right]
+O(\lambda^2).
}
\label{eq:app_dq_mod}
\end{equation}
This term combines the curvature of the transformed residuals with the
parameter-dependent Gaussian-volume contribution discussed in
\cref{sec:covariance}. The remainder \(O(\lambda^2)\) denotes terms of
second and higher order for a generic value of \(p\); it does not imply
that a nonzero \(O(\lambda^2)\) contribution remains when the leading
term happens to vanish. In particular, at \(p=1\) the Box--Cox
transformation reduces to \(h_1(x)=x-1\), which is only a constant
translation of the original variable. The model-centered Gaussian
reconstruction then coincides exactly with the original marginal model,
so \(q_{{\rm mod},1}=q_E\) to all orders in \(\lambda\).

For the data-centered prescription one instead obtains
\begin{equation}
\boxed{
q_{{\rm dat},p}-q_E
=
-\lambda Z
\left[
2\beta(Z^2+H-1)
+
\kappa\{Z^2+(2+\alpha)H\}
\right]
+O(\lambda^2).
}
\label{eq:app_dq_dat}
\end{equation}
Here the first term,
\begin{equation}
-2\beta\lambda Z(Z^2+H-1),
\end{equation}
is the data-dependent-metric contribution already present in the
PPP problem, while the term proportional to $\kappa=p-1$
is generated by the nonlinear representation.

The connection with the ordering coefficients in \cref{sec:ordering} is
most transparent at the Gaussian-limit $68.27\%$ boundaries.  At zeroth
order these are $Z=\pm1$.  If the approximate and exact LR statistics differ
by a first-order term, the corresponding first-order displacement of a
boundary is obtained by dividing that difference by the local derivative
of $Z^2$, namely $2Z$.

For the model-centered prescription,
\cref{eq:app_dq_mod} therefore gives
\begin{equation}
Z_A^\pm-Z_E^\pm
=
-\frac{\kappa}{2}
\left[
-3+(2+\alpha)H
\right]\lambda
+O(\lambda^2).
\label{eq:app_boundary_mod}
\end{equation}
The two edges move by the same amount.  At fixed $H$ the leading probability
exchanged between the exact and approximate acceptance regions is thus
\begin{equation}
\Dord^{\rm mod}(H)
=
|\kappa|\,\phione\,
\left|-3+(2+\alpha)H\right|\lambda
+O(\lambda^2).
\end{equation}
Averaging over $H\sim\chi_1^2$ gives
\begin{equation}
a_1^{\rm mod}(\alpha,p)
=
|p-1|\phione\,
\E\left|-3+(2+\alpha)H\right|,
\end{equation}
which is \cref{eq:a1_mod}.

For the data-centered prescription,
\cref{eq:app_dq_dat} similarly gives
\begin{equation}
Z_A^\pm-Z_E^\pm
=
\left\{
\beta H
+
\frac{\kappa}{2}
\left[1+(2+\alpha)H\right]
\right\}\lambda
+O(\lambda^2).
\label{eq:app_boundary_dat}
\end{equation}
The corresponding exchanged probability, averaged over the ancillary
distribution, is
\begin{equation}
a_1^{\rm dat}(\alpha,p)
=
2\phione\,
\E\left|
\frac{p-1}{2}
+
\left[
1-\alpha+\frac{p-1}{2}(2+\alpha)
\right]H
\right|,
\end{equation}
recovering \cref{eq:a1_dat}.

Away from zeros of the first-order boundary displacement,
\begin{equation}
\Dord=a_1\lambda+a_3\lambda^3+O(\lambda^5).
\end{equation}
The PPP construction is a special case because its first-order displacement,
\(d_1(H)=\beta H\), vanishes at the endpoint \(H=0\).  In the shrinking
region \(H=O(\lambda)\), the nominally second-order boundary displacement
therefore competes with the first-order term, so the fixed-\(H\) expansion
is not uniform.  Since \(H\sim\chi_1^2\) has density
\(f_H(H)\propto H^{-1/2}\) near the origin, this boundary layer contains
\(O(\lambda^{1/2})\) probability.  Multiplying this by the
\(O(\lambda^2)\) boundary displacement produces an exceptional
\(O(\lambda^{5/2})\) contribution to the averaged ordering distance.
Thus, for the PPP construction, a fractional-power correction appears before the
regular \(O(\lambda^3)\) term, while the leading \(O(\lambda)\) result is
unchanged.

More generally, for \(H\sim\chi^2_\nu\) with \(\nu=N-1\), the probability
contained in the boundary layer \(H=O(\lambda)\) scales as
\(O(\lambda^{\nu/2})\), so its contribution to the ordering distance is
\(O(\lambda^{2+\nu/2})=O(\lambda^{(N+3)/2})\).  Thus the anomalously early
\(O(\lambda^{5/2})\) correction is specific to \(N=2\); for \(N=3\) it is
of order \(O(\lambda^3)\), and for larger \(N\) it is still more suppressed.

\paragraph{Arbitrary-$N$ PPP ordering.}
The preceding transformed-likelihood expansion refers to the
two-measurement geometry.  For the untransformed data-centered prescription,
$p=1$, the reduction of \cref{eq:app_general_plugin_Q} allows the
likelihood-ratio calculation to be performed for arbitrary $N$.

Let
\begin{equation}
\nu=N-1,
\qquad
H\sim\chi^2_\nu,
\qquad
Z\perp H.
\end{equation}
The exact marginal likelihood-ratio statistic remains
\begin{equation}
q_E
=
Z^2
-2\beta\lambda Z
+
\beta\left[(2\beta-1)Z^2+\beta\right]\lambda^2
+O(\lambda^3).
\label{eq:app_qE_N}
\end{equation}
For the data-centered PPP construction one obtains
\begin{align}
q_{\rm PPP}
={}&
Z^2
-2\beta\lambda Z(Z^2+H)
\nonumber\\
&+
\beta\lambda^2
\left[
(4\beta-1)Z^4
+(5\beta-2)HZ^2
+\beta H^2
\right]
+O(\lambda^3).
\label{eq:app_qPPP_N}
\end{align}
Consequently,
\begin{equation}
\boxed{
q_{\rm PPP}-q_E
=
-2\beta\lambda Z
\left(Z^2+H-1\right)
+O(\lambda^2).
}
\label{eq:app_dqPPP_N}
\end{equation}

At the Gaussian-limit boundary $q=1$, write the first-order boundary
coefficient of a given construction as
\begin{equation}
Z_\pm=\pm1+b_1(H)\lambda+O(\lambda^2).
\end{equation}
For the exact and PPP constructions,
\begin{equation}
b_1^E=\beta,
\qquad
b_1^{\rm PPP}(H)=\beta(1+H).
\end{equation}
We reserve $d_1(H)$ for the relative approximate-minus-exact displacement
coefficient used in the main text:
\begin{equation}
d_1(H)\equiv b_1^{\rm PPP}(H)-b_1^E=\beta H.
\end{equation}
Thus the relative displacement of the two equally calibrated acceptance
intervals is, to first order,
\begin{equation}
\boxed{
\delta Z(H)=d_1(H)\lambda=\beta H\lambda+O(\lambda^2).
}
\label{eq:app_deltaZ_N}
\end{equation}

At fixed $H$ the corresponding leading exchanged probability is
\begin{equation}
\Dord(H)
=
2\phione\,\beta H\lambda
+\text{higher-order terms}.
\end{equation}
Using $\E[H]=\nu=N-1$ therefore gives
\begin{equation}
\boxed{
\Dord
=
2\phione\,\beta(N-1)\lambda
+\text{higher-order terms}.
}
\label{eq:app_Dord_N}
\end{equation}

\section{Second-order coverage coefficients}
\label{app:coverage}

Write
\begin{equation}
q=Z^2+\lambda q_1+\lambda^2q_2+\cdots,
\end{equation}
and
\begin{equation}
c_\gamma=1+g_2\lambda^2+O(\lambda^4).
\end{equation}
Writing the positive boundary of the construction under consideration as
$Z_+=1+b_1(H)\lambda+O(\lambda^2)$, expansion of the two boundaries gives
\begin{equation}
g_2=
\E_H\left[
2b_1(H)^2+b_1(H)q_1'(1,H)+q_2(1,H)
\right].
\end{equation}
Consequently,
\begin{equation}
\Pp(q\le1)
=
\gamma-\phione g_2\lambda^2+O(\lambda^4).
\end{equation}

For the exact likelihood,
\begin{equation}
g_2^E=3\alpha^2-5\alpha+2.
\end{equation}
Relative to the exact likelihood, the five approximate $q\le1$ coverage
coefficients are
\begin{center}
\begin{tabular}{lc}
\toprule
Approximation &
$\Dfix/(\phione\lambda^2)$\\
\midrule
model-centered, $p=1/2$ & $(5\alpha^2-54\alpha+11)/16$\\
model-centered, log & $(3\alpha^2-90\alpha-23)/12$\\
data-centered, $p=1$ & $1-\alpha^2$\\
data-centered, $p=1/2$ & $(35+18\alpha-43\alpha^2)/16$\\
data-centered, log & $(25+6\alpha-57\alpha^2)/12$\\
\bottomrule
\end{tabular}
\end{center}

\paragraph{Arbitrary-$N$ PPP coverage.}

Using the arbitrary-$N$ PPP statistic of
\cref{eq:app_qPPP_N}, the first-order boundary coefficient is
\begin{equation}
b_1^{\rm PPP}(H)=\beta(1+H).
\end{equation}
Substitution into the integrand of the general second-order expression above
gives the $H$-dependent quantity
\begin{equation}
\widetilde g_2^{\rm PPP}(H)
=
\beta
\left[
\beta H^2+\beta H-2H-1
\right].
\label{eq:app_g2PPP_N}
\end{equation}
For the exact likelihood,
\begin{equation}
g_2^E
=
\beta(3\beta-1),
\end{equation}
which is equivalent to
$3\alpha^2-5\alpha+2$.

With
\begin{equation}
H\sim\chi^2_\nu,
\qquad
\nu=N-1,
\end{equation}
the required moments are
\begin{equation}
\E[H]=\nu,
\qquad
\E[H^2]=\nu(\nu+2).
\end{equation}
Hence the scalar critical-value coefficient is
\begin{equation}
g_2^{\rm PPP}
=
\E[\widetilde g_2^{\rm PPP}(H)]
=
\beta
\left[
\beta\nu(\nu+3)-2\nu-1
\right].
\end{equation}
Taking the difference between the approximate and exact fixed-threshold
coverages gives
\begin{equation}
\boxed{
\frac{\Dfix}{\phione\lambda^2}
=
\beta
\left[
2\nu+3\beta-\beta\nu(\nu+3)
\right]
+O(\lambda^2),
\qquad
\nu=N-1.
}
\label{eq:app_Dfix_N}
\end{equation}

For $N=2$, $\nu=1$, this reduces to
\begin{equation}
\frac{\Dfix}{\phione\lambda^2}
=
\beta(2-\beta)
=
1-\alpha^2,
\end{equation}
in agreement with the data-centered $p=1$ entry in the table above.

Besides the trivial $\beta=0$ solution,
\cref{eq:app_Dfix_N} vanishes at
\begin{equation}
\boxed{
\beta_{\rm blind}(N)
=
\frac{2\nu}
{\nu(\nu+3)-3},
\qquad
\nu=N-1.
}
\label{eq:app_beta_blind_N}
\end{equation}
For $N=2$ this gives the unphysical value $\beta=2$.
For every $N\ge3$, however, the solution lies in the physical interval
$0<\beta<1$.  In particular,
\begin{equation}
N=3:
\qquad
\beta_{\rm blind}=\frac47,
\qquad
\alpha_{\rm blind}=\frac37,
\end{equation}
while for $N=4$,
\begin{equation}
\beta_{\rm blind}=\frac25,
\qquad
\alpha_{\rm blind}=\frac35.
\end{equation}

\section{Recognizable tail subsets}
\label{app:tails}

For a fixed threshold \(h_0>0\), independent of \(\lambda\), let
\(\mathcal S(h_0)=\{H>h_0\}\) and define the conditional coverage
\(C[\mathcal S(h_0)]\equiv
\Pp_\mu(A_A\mid\mathcal S(h_0))\).  Because this subset excludes the
endpoint \(H=0\), the boundary-layer contribution discussed above is
absent, and the regular perturbative expansion can be averaged over the
conditional ancillary distribution:
\begin{align}
\dcond[\mathcal S(h_0)]
&=
2\phione\lambda^2
\E[e(H)\mid\mathcal S(h_0)]+O(\lambda^4),
\\
\Dord[\mathcal S(h_0)]
&=
2\phione\lambda
\E[|d_1(H)|\mid\mathcal S(h_0)]+O(\lambda^3).
\end{align}
With $H=U^2$ and $U\sim \mathcal N(0,1)$, the required tail moments are
\begin{align}
\Pp(H>h_0)&=2\bar\Phi(\sqrt{h_0}),
\\
\E[H\mid\mathcal S(h_0)]
&=
1+
\frac{\sqrt{h_0}\,\phi(\sqrt{h_0})}
     {\bar\Phi(\sqrt{h_0})},
\\
\E[H^2\mid\mathcal S(h_0)]
&=
3+
\frac{(h_0^{3/2}+3\sqrt{h_0})\phi(\sqrt{h_0})}
     {\bar\Phi(\sqrt{h_0})}.
\end{align}
where \(\bar\Phi(z)\equiv1-\Phi(z)\).

\section{Buehler relevant-subset reconnaissance}
\label{app:buehler}

For the PPP construction, denote by $A_{\rm PPP}$ the calibrated acceptance region of
the data-centered construction at the parameter value under
test.  For a fixed ancillary subset $\mathcal S$, define
\begin{equation}
M_1=\E[H\mid\mathcal S],
\qquad
M_2=\E[H^2\mid\mathcal S].
\end{equation}
Returning from $(\lambda,\alpha)$ to the physical model with fixed $s$ and
$\tau$, the leading conditional-coverage defect can be written
\begin{equation}
\Pp_\mu(A_{\rm PPP}\mid\mathcal S)-\gamma
=
\phione s^2
\left[
2(M_1-1)+
\beta(\mu)(4-M_1-M_2)
\right]
+O(s^4),
\end{equation}
where
\begin{equation}
\beta(\mu)=
\frac{s^2\mu^2}{s^2\mu^2+\tau^2}
\in(0,1).
\end{equation}
Since this is linear in $\beta(\mu)$, a sufficient leading-order condition for
uniform negative conditional bias is
\begin{equation}
M_1<1,
\qquad
M_2>M_1+2.
\end{equation}
A simple one-sided high-$H$ subset fails the first condition, while a low-$H$
subset generally fails the second.  Disconnected subsets combining a narrow
very-low-$H$ region with a rare high-$H$ tail can instead satisfy both
conditions.

To test whether this perturbative possibility survives at finite normalization
uncertainty, we choose one such subset once and keep it fixed throughout the
numerical study,
\begin{equation}
\mathcal S_\star
=
\{H<0.0586\}\cup\{H>6.7457\}.
\label{eq:buehler_subset}
\end{equation}
The thresholds were selected from the leading-order criterion by maximizing
the smaller of the two endpoint undercoverage margins, subject to
$\Pp(\mathcal S_\star)\ge 0.20$.  For this choice,
\begin{equation}
\Pp(\mathcal S_\star)=0.2007,
\qquad
M_1=0.4193,
\qquad
M_2=3.5911.
\end{equation}
Thus the coefficient of the leading conditional-coverage defect is negative
throughout the physical interval $0<\beta<1$.

We then evaluate the full finite-$s$ PPP likelihood-ratio construction rather
than its perturbative expansion.  For fixed $s$ and $\tau$, varying the
parameter $\mu>0$ is equivalent to scanning
\begin{equation}
\beta(\mu)
=
\frac{s^2\mu^2}{s^2\mu^2+\tau^2}
\in(0,1).
\end{equation}
At each value of $\beta$ the PPP acceptance region is calibrated
unconditionally to the nominal coverage $\gamma$, after which we evaluate the
conditional coverage
\begin{equation}
\Pp_\mu(A_{\rm PPP}\mid\mathcal S_\star)-\gamma .
\end{equation}
The integration over the informative Gaussian direction is performed
continuously by solving the likelihood-ratio boundaries, while the remaining
ancillary integration is evaluated by deterministic quadrature.

\begin{figure}[t]
\centering
\includegraphics[width=0.88\linewidth]{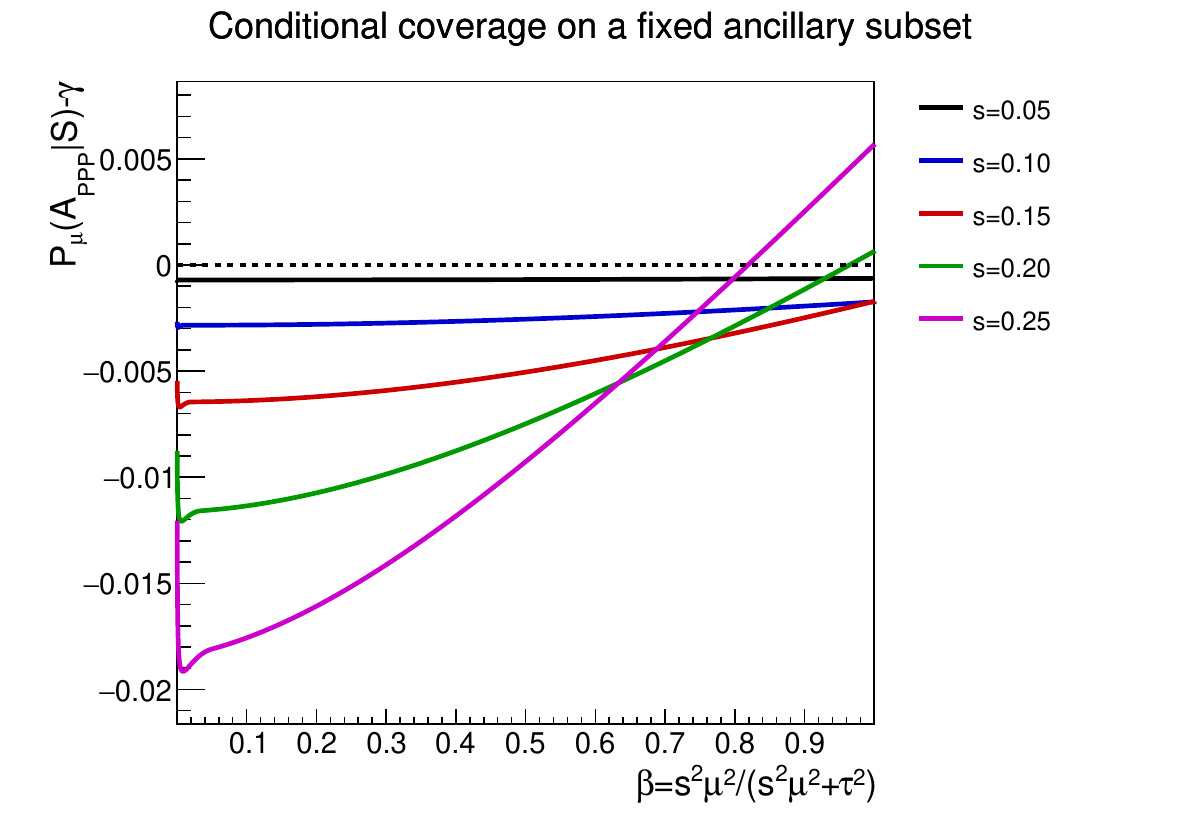}
\caption{
Finite-$s$ reconnaissance of a negatively biased relevant subset for the
$p=1$ PPP construction.
The fixed ancillary subset is
$\mathcal S_\star=\{H<0.0586\}\cup\{H>6.7457\}$, with
$\Pp(\mathcal S_\star)=0.2007$, chosen once from the leading-order conditions
of Appendix~F and then held fixed throughout the scan.
The curves show the conditional coverage defect
$\Pp_\mu(A_{\rm PPP}\mid\mathcal S_\star)-\gamma$
as a function of
$\beta=s^2\mu^2/(s^2\mu^2+\tau^2)$, which spans the positive-$\mu$ parameter
range at fixed $s$ and $\tau$.
For $s=0.05$, $0.10$, and $0.15$ the defect remains negative throughout the
numerical scan, providing finite-$s$ evidence for a negatively biased relevant
subset.
For $s=0.20$ and $0.25$ the defect eventually becomes positive toward the
upper part of the $\beta$ range, showing that the property does not persist to
arbitrarily large normalization uncertainty.
The belts are separately calibrated to the nominal $68.27\%$ unconditional
coverage at every parameter point.
}
\label{fig:buehler_scan}
\end{figure}

The result is shown in \cref{fig:buehler_scan}.  For
\(s=0.05\), \(0.10\), and \(0.15\), the conditional coverage defect remains
negative throughout the numerical scan of the physical parameter range.  At
\(s=0.20\) and \(0.25\) it eventually becomes positive toward the upper end of
the \(\beta\) range.

The calculation therefore provides finite-\(s\) numerical evidence that the
fixed subset remains negatively biased throughout the sampled parameter range
for \(s\leq0.15\).  This is consistent with a negatively biased relevant
subset, but the finite scan is not by itself a proof of the uniform statement
over the continuum \(0<\beta<1\).  At larger normalization uncertainty the
property need not persist, as illustrated by the sign change observed for
\(s=0.20\) and \(0.25\).  We regard this as supporting evidence for the
ancillary-leakage mechanism rather than as a necessary ingredient of the main
result.


\section{Numerical validation}
\label{app:numerics}

Numerical checks of the main results in this manuscript were performed with ROOT/C++ macros using continuous solution of the LR boundaries in $Z$ and deterministic quadrature over the ancillary direction, thus avoiding the probability quantization that appears if the acceptance region is evaluated on a binary $(Z,U)$ grid.

For the two-measurement PPP benchmark introduced in
\cref{s:3descr}, with
\(\sigma_1=\sigma_2=0.5\), \(s=0.2\), and
\(\mu_0=10.5\), one has
\begin{equation}
\lambda=0.202815,\qquad
\alpha=0.0275634.
\end{equation}
The exact calibrated critical value is
\begin{equation}
c_{68}=1.07445.
\end{equation}
A $10^6$-toy interval inversion gives decision-swap probabilities
\begin{center}
\begin{tabular}{lc}
\toprule
Method & swap probability\\
\midrule
data-centered, $p=1$ & $8.668\%$\\
data-centered, $p=1/2$ & $3.644\%$\\
data-centered, log & $4.979\%$\\
\bottomrule
\end{tabular}
\end{center}
in agreement with the direct acceptance-region $\Dord$ calculation.

For comparison, at the two-measurement data-centered logarithmic
coverage-blind point of \cref{sec:blind},
\[
\alpha=0.716986,\qquad \lambda=0.1,
\]
the corresponding decision-swap probabilities for the data-centered
\(p=1\), \(p=\tfrac12\), and \(p=0\) constructions are
\begin{equation}
1.372\%,\qquad 3.218\%,\qquad 7.652\%,
\end{equation}
respectively.

\paragraph{Arbitrary-$N$ PPP validation.}
The arbitrary-$N$ expressions of
\cref{eq:app_Dord_N,eq:app_Dfix_N} were checked independently using
deterministic integration over
\begin{equation}
Z\sim\mathcal N(0,1),
\qquad
H\sim\chi^2_{N-1},
\end{equation}
with the exact finite-$\lambda$ likelihood-ratio statistics and separately
calibrated exact and approximate critical values.

In the following comparisons the superscript ``LO'' denotes the leading
perturbative approximation, $O(\lambda)$ for $\Dord$ and $O(\lambda^2)$ for
$\Dfix$.
For the $N=2$ cross-check with
$\lambda=0.05$ and $\beta=0.5$, we obtain
\begin{equation}
\Dord=0.0120615,
\qquad
\Dord^{\rm LO}=0.0120985,
\end{equation}
while
\begin{equation}
\Dfix=4.526\times10^{-4},
\qquad
\Dfix^{\rm LO}=4.537\times10^{-4}.
\end{equation}
This reproduces the previously derived two-measurement result.

At the new $N=3$ leading coverage-blind point,
\begin{equation}
\lambda=0.1,
\qquad
\beta=\frac47,
\qquad
\alpha=\frac37,
\end{equation}
the separately calibrated critical values are
\begin{equation}
c_E=1.0040508,
\qquad
c_A=1.0040220.
\end{equation}
The exact finite-$\lambda$ ordering discrepancy is
\begin{equation}
\Dord=0.05390,
\end{equation}
compared with the leading prediction
\begin{equation}
\Dord^{\rm LO}=0.05531.
\end{equation}
At the same point,
\begin{equation}
\Dfix=1.71\times10^{-5},
\end{equation}
while its $O(\lambda^2)$ coefficient vanishes identically.

For $N=4$, at the corresponding blind point
$\beta=0.4$ and $\lambda=0.1$, we find
\begin{equation}
\Dord=0.05682,
\qquad
\Dord^{\rm LO}=0.05807,
\end{equation}
and
\begin{equation}
\Dfix=-1.23\times10^{-5}.
\end{equation}

As a separate check of the reduction to $(m,H)$ for unequal statistical
errors, a $10^6$-toy $N=3$ simulation used relative error scales
\begin{equation}
\sigma_1:\sigma_2:\sigma_3=1:1.7:2.4,
\end{equation}
rescaled to the same total statistical precision.  It gave
\begin{equation}
\E[H]=1.9993,
\qquad
\Var(H)=3.9941,
\end{equation}
in agreement with the $\chi^2_2$ expectations $2$ and $4$.
Most importantly, evaluation of the PPP likelihood ratio in the full
three-dimensional measurement space and through the reduced $(m,H)$
expression agreed event by event to
\begin{equation}
\max
\left|
q_{\rm PPP}^{\rm full}
-
q_{\rm PPP}^{(m,H)}
\right|
=
3.6\times10^{-14}.
\end{equation}
This verifies numerically that the arbitrary heteroscedastic problem reduces
to the same two statistics to machine precision.

A small number of toys fall outside the positive domain required by log and
square-root transformations because the Gaussian model has formally
unbounded support.  In the transformed-variable interval studies, toys with
$x_1\leq0$ or $x_2\leq0$ are excluded before interval inversion.  Their
fraction in the million-toy studies is below $10^{-4}$, so this domain
restriction has no visible effect on the quoted results.

\end{appendices}


\bibliographystyle{unsrt}
\bibliography{references}

\end{document}